\documentstyle[emulate_apj,apjfonts,epsf]{article}

\gdef\h50min{$h_{50}^{-1}$}
\lefthead{van Dokkum et al.}
\righthead{CM relation in CL\,1358+62}
\slugcomment{Accepted for publication in the Astrophysical Journal}
\begin{document}
\title{The color-magnitude relation in CL\,1358+62 at $z=0.33$:
evidence for significant evolution in the S0 population}
\author{Pieter G. van Dokkum, Marijn Franx}

\affil{Kapteyn Astronomical Institute, P.O. Box 800, NL-9700 AV,
Groningen, The Netherlands}

\author{Daniel D.  Kelson, Garth D. Illingworth}

\affil{University of California Observatories / Lick Observatory, 
Board of Studies in Astronomy and Astrophysics, 
University of California, Santa Cruz, CA 95064}

\author{David Fisher}
\affil{Kapteyn Astronomical Institute, P.O. Box 800, NL-9700 AV,
Groningen, The Netherlands}

\and

\author{Daniel Fabricant}
\affil{Harvard-Smithsonian Center for
Astrophysics, 60 Garden Street, Cambridge, MA \ 02318}

\begin{abstract}

We use a large, multi-color mosaic of HST WFPC2 images to measure the
colors and morphologies of 194 spectroscopically confirmed members of
the rich galaxy cluster CL\,1358+62 at $z=0.33$.  We study the
color-magnitude (CM) relation as a function of radius in the cluster,
to a limit of 4.6 arcmin from the center, equivalent to 1.6 \h50min
Mpc.

The intrinsic scatter in the restframe $B-V$ CM relation of the
elliptical galaxies is very small: $\sim$0.022 magnitudes.  The CM
relation of the ellipticals does not depend significantly on the
distance from the cluster center.  In contrast, the CM relation for
the S0 galaxies does depend on radius: the S0s in the core follow a CM
relation similar to the ellipticals, but at large radii ($R >
0.7$\,\h50min\,Mpc) the S0s are systematically bluer and the scatter
in the CM relation approximately doubles to $\sim$0.043
magnitudes. The blueing of the S0s at large radii is significant at
the $95$\,\% confidence level.

These results imply that the S0 galaxies in the outer parts of the
cluster have formed stars more recently than the S0s in the inner
parts.  A likely explanation is that clusters at $z=0.33$ continue to
accrete galaxies and groups from the field and that infall
extinguishes star formation.  The apparent homogeneity of the
elliptical galaxy population implies that star formation in recently
accreted ellipticals was terminated well before accretion occurred.

We have constructed models to explore the constraints that these
observations place on the star formation history of cluster galaxies.
The best constrained parameter is the scatter in the
luminosity-weighted age $\Delta \tau_L / \langle \tau_L \rangle$,
which is less than 18\,\% for the ellipticals and the S0s in the
cluster core, and less than 35\,\% for the S0s in the outer parts of
the cluster. The constraints on the most recent period of star
formation are model dependent, but we show that star formation in
ellipticals likely ceased at $z=0.6$ or higher.  If we assume that the
galaxies have a constant
star formation rate up to a randomly distributed truncation
time, we find that the S0s in the outer parts of the cluster have
experienced star formation until the epoch of observation at $z=0.33$.

We conclude that the population of S0s in clusters is likely to evolve
as star forming galaxies are converted into passively evolving
galaxies.  Assuming a constant accretion rate after $z=0.33$, we
estimate $\sim$15\% of the present day early-type galaxy population
in rich clusters was accreted between $z=0.33$ and $z=0$.  The
ellipticals (and the brightest S0s) are probably a more stable
population, at least since $z=0.6$.

\end{abstract}


\section{Introduction}

Early-type galaxies in nearby rich clusters form a homogeneous
population. The elliptical and S0 galaxies in clusters such as Coma
follow a tight color-magnitude (CM) relation (Bower, Lucey, \& Ellis
1992b), Mg$_2$--$\sigma$ relation (Guzman et al.\ 1992), and
Fundamental Plane (e.g., J\o{}rgensen, Franx, \& Kj\ae{}rgaard 1993).
The low scatter in these relations implies that the spread in the
metallicities and ages of the galaxies is small at a given mass or
luminosity.

The homogeneity of the early-type galaxies constrains models for the
formation and evolution of these galaxies (e.g., Sandage \&
Visvanathan 1978, Faber et al.\ 1987, Bower et al.\ 1992b).  In
particular, the small intrinsic scatter in the CM relations of Coma
and Virgo implies that the early-type galaxies in these clusters
either formed at high redshift, or that their formation was
synchronized (Bower et al.\ 1992b).

If the intrinsic scatter is due to age differences between the
galaxies, one expects the scatter to be higher at larger lookback times.
Hence, the scatter in the CM relation is expected to increase with redshift.
Ellis et al.\ (1997) studied the CM relation in three clusters at
$z \sim 0.55$ using Hubble Space Telescope (HST) data.  Surprisingly,
Ellis et al.\ (1997) found that the CM relation of the ellipticals and
S0s is still very tight at $z=0.55$.  This led these authors to
conclude that the star formation in early-type galaxies likely ceased
at much higher redshifts.  A similar conclusion was reached by
Andreon, Davoust, \& Heim (1997), who found a tight CM relation of the
early-type galaxies in Abell 851 at $z=0.4$.  These results were
further strengthened by Stanford, Eisenhardt, \& Dickinson (1997), who found
that the scatter in the CM relation of early-type galaxies in rich
clusters is nearly constant with redshift to $z \sim 0.9$.
Similarly, studies of the Fundamental Plane relation in clusters to $z
\sim 0.6$ show that the scatter in the relation is not very
different from that in local clusters (van Dokkum \& Franx 1996;
Kelson et al.\ 1997; Bender et al.\ 1997), indicating that the population of early-type
galaxies has been stable and homogeneous over a significant fraction
of the age of the universe.

The result of Ellis et al.\ (1997) is surprising, because it seems
difficult to reconcile with the strong evolution of the cluster
population implied by the Butcher-Oemler effect.  The enhanced
fraction of blue galaxies in intermediate redshift clusters as
observed by Butcher \& Oemler (1978, 1984) is often interpreted as the
result of the transformation of blue field galaxies to red cluster
galaxies (e.g., Butcher \& Oemler 1984, Abraham et al.\ 1996b).  A key
question is whether the end products of this transformation are
early-types (e.g., Dressler et al.\ 1997), or whether the
Butcher-Oemler effect can be largely explained by blue spirals turning
red (Butcher \& Oemler 1984).  Many blue galaxies in intermediate
redshift clusters have the characteristics of normal (field) spirals
(e.g., Couch \& Sharples 1987, Andreon et al.\ 1997).  However, some
have a starburst or a post starburst spectrum (Dressler \& Gunn 1983;
Couch \& Sharples 1987), and it seems natural to link these post
starburst galaxies to today's early-type galaxies. If a population of
recently transformed early-types exists, they are expected to be bluer
than the pre-existing early-types, and to increase the scatter in the
CM relation.  Apparently, such galaxies are not in the samples studied
by Ellis et al.\ (1997) and Stanford et al.\ (1997).
On the other hand, the strong evolution
with redshift of the morphology-density relation claimed by
Dressler et al.\ (1997) is in qualitative agreement with the continuous
transformation of spirals to S0s.

The studies of Ellis et al.\ (1997) and Stanford et al.\ (1997) suffer
from two limitations: membership information is sparse, and the field
sizes are small.  The membership information is important because it
is crucial to measure the full distribution of galaxies in the
color-magnitude plane, and the blue side of the CM relation is heavily
contaminated by field galaxies.  Usually, a correction is applied by
subtracting the expected number of field galaxies based on number
counts. Since the number counts can vary in a small field, it is
obviously preferable to have direct spectroscopic membership
information.

A large angular coverage is of particular importance, since there is
good evidence that the blue galaxies are more abundant at larger
distances from the cluster center (e.g., Butcher \& Oemler 1984,
Pickles \& van der Kruit 1991, Abraham et al. 1996b). An environmental
dependence is also indicated by the evidence for young populations in
early-type galaxies in the field (e.g., Larson, Tinsley, \& Caldwell
1980, Bothun \& Gregg 1990).  If there are (mildly) blue early-type
galaxies in intermediate redshift clusters, it seems likely that they
reside in the transition region between the cluster and the field.

In the present study, we extend the study of the CM relation to larger
radii, using a large $8' \times 8'$ HST mosaic of the cluster
CL\,1358+62 at $z=0.33$.  We determine morphologies, colors, and
magnitudes for 194 spectroscopically confirmed cluster members within
the HST mosaic.  The color-magnitude relation in intermediate redshift
clusters has been studied extensively (e.g., Aragon-Salamanca et al.\
1993, Rakos \& Schombert 1995, Abraham et al.\ 1996b, Ellis et al.\
1997, Stanford et al.\ 1997), but never before with high resolution,
large format images.  The aims of this study are to constrain the star
formation histories of the galaxies, and to establish whether the
histories depend on morphology or environment.  In particular, we test
whether star formation in the ellipticals and S0 galaxies in the outer
parts of the cluster has persisted to more recent epochs than star
formation in early-type galaxies in the cluster core.

\section{Data}

\subsection{Sample Selection}

The present sample is based on a large spectroscopic survey of the
cluster. The spectra were obtained with multislit masks at the
Multiple Mirror Telescope and the William Herschel Telescope.  The
sample selection and the reduction and analysis of the spectroscopic
data are described in detail in Fabricant, McClintock, \& Bautz
(1991), and Fisher et al.~(1997). Here, we briefly summarize the
sample selection.

The spectroscopic sample was selected on the basis
of $R$ magnitude. Twenty multislit aperture masks were exposed within
a $\sim$10$^{\prime}$$\times$11$^{\prime}$ field centered on the
brightest cluster galaxy in CL\,1358+62, and redshifts for 387
galaxies were determined. The sample of galaxies with redshifts is
$>80$\,\% complete to $R = 21$, or $F814W \sim 20.2$, dropping steeply
to $\sim 20$\,\% at $R = 22$ ($F814W \sim 21.2$). The completeness within
the $8' \times 8'$ HST WFPC2 field (see below) is somewhat higher:
$>90$\,\% to $R=21$, and $\sim 30$\,\% at $R=22$.
This incompleteness is caused by the limited number of galaxies that was
observed, and not by the inability to measure the redshifts of observed
galaxies. The ``success rate'' of measuring the redshift is $100$\,\%
to $R=21.1$. The success rate drops to $70$\,\% at $R=22.3$
($F814W \sim 21.5$), and $50$\,\% at $R=23$.
The success rate does not depend strongly on color,
for $R<23.5$ (Fisher et al.\ 1997).

We refer the reader
to Fisher et al.~(1997) for a discussion of cluster membership and
substructure in the cluster.  Following Fisher et al., we include as
cluster members all galaxies in the interval $0.31461 < z < 0.34201$.
The subsample used for this paper consists of all 194 confirmed
cluster members in the $8' \times 8'$ HST WFPC2 field (see below).
For the analysis in Sect.~\ref{cm.sec},
we furthermore impose a magnitude limit of
$F814W = 21.5$, reducing the sample to 188 galaxies.

\subsection{Imaging}

WFPC2 images from twelve HST pointings were combined to give an $\sim
8' \times 8'$ field centered on the brightest cluster galaxy (BCG) in
CL\,1358+62. The cluster was observed with the $F606W$ and $F814W$
filters (close to rest frame $B$ and $V$), on February 15, 1996.
Exposure times were 3600\,s in each filter, for each pointing.  Figure
\ref{mosaic.plot} [Plate 1] shows the field layout.

\subsubsection{Reduction}

The pipeline reduction was performed at the Space Telescope Science
Institute (STScI). No recalibration was done, and we verified that the
most recent calibration files had been used in the pipeline
processing.  The crucial step in the data reduction was the sky
subtraction.  The cluster was observed in the Continuous Viewing Zone
(CVZ) of HST.  The sky background is fairly high in most exposures,
due to reflected Earth light.  The Earth light reflects off the Optical
Telescope Assembly (OTA) and scatters throughout the WFPC2 field.
In most exposures, dark diagonal bands are present; these
appear when the OTA vanes pass into the shadow of the WFPC2 camera
relay spiders (see Biretta, Ritchie, \& Rudloff ~1995).

Three identical 1200\,s exposures were obtained in each filter.  The
exposures were timed to minimize the scattered light in one of the
three exposures. For the majority of exposures, we modelled the
background in two steps. First, the image with the lowest background
was subtracted from the two other images. Second, the residual images
(containing the background structure + noise) were fitted with high
order 2D polynomials.  These fits were then subtracted from the high
background images.  Finally, we checked the results by eye
and iterated to minimize the residuals of the cross patterns.  In all
$F606W$ exposures this method worked well.  However, in four of the
twelve $F814W$ pointings the background structure was present in all
three exposures. For these pointings, the cross pattern in the lowest
background image was carefully fitted and subtracted before
subtracting this image from the higher background images.

In 10 of the 72 exposures bright linear features are present, due to
the passage of bright objects through the field of view during the
exposures (Biretta et al.~1995).  The areas affected by these features
were masked before the three exposures of each field were combined.
When both a high background and a bright streak are present, the
bright streak was masked before modeling the background.

The individual exposures were combined with the {\sc crreject}
task, incorporated in the {\sc iraf} {\sc stsdas} reduction
package. The output cosmic ray mask files were visually checked to verify
that only cosmic rays were removed.  Hot pixels were corrected by
creating a ``dark'' frame from the exposures themselves; the twelve
pointings were median filtered to identify the locations of hot
pixels. Bad pixels and bad columns were removed by interpolation.

The full field is shown in Fig. \ref{mosaic.plot} [Plate 1].  The
total area of the image is 49\,arcmin$^2$.  The galaxy density
decreases with distance from the BCG, allowing us to study the
dependence of the properties of the cluster galaxies on local galaxy
density.  Fig. \ref{center.plot} [Plate 2] is a color image of the
central part of the cluster.  Clearly, most of the cluster galaxies
have very similar colors, but our spectroscopy shows that many of the
blue galaxies are also cluster members (Fisher et al. 1997).



\subsubsection{Zeropoints}
The $F606W$ and $F814W$ HST WFPC2 filters are close to the rest frame
$B$ and $V$ bands for an object with $z=0.33$.  As noted by many
authors, standard $K$-corrections may introduce large errors, due to
the effects of spectral evolution and intrinsic scatter in galaxy
colors.  Therefore, we followed the procedure described in van Dokkum
\& Franx (1996), and derived direct transformations from the HST
filters to the redshifted $V$ band, denoted with $V_z$, and $(B-V)_z$
colors:
\begin{equation}
V_z = F814W + 0.20 (F606W - F814W) + 0.65
\end{equation}
\begin{equation}
(B-V)_z = 0.82 (F606W - F814W) - 0.13
\end{equation}
For the early-type galaxies in CL\,1358+62, $V_z \sim F814W + 0.9$.
The zero points of the $F814W$ and $F606W$ filters are based on the
Vega spectrum, and taken from Table 41.1 in the HST Data Handbook
(Leitherer 1995).  Using spectral energy distributions from
Pence~(1976), we find the transformations are independent of galaxy
type to $\sim 0.01$ magnitudes, for early-types and early-type
spirals.

\subsection{Morphologies and Photometry}
\subsubsection{Visual Classification}
\label{visclass.sec}

We have extracted $6\farcs4 \times 6\farcs4$ images of all 194
galaxies from the WFPC2 frames. Greyscale representations of these
images are presented in Fig.~\ref{allgal.plot} [Plate 3].  Four of us
(PvD, MF, DK, and GDI) morphologically classified the galaxies.  Our
classification scheme is identical to that used by Dressler (1980) for
galaxies in nearby clusters.  Five morphological types are
distinguished: E, S0, S, Irregular, and unclassified.  These
morphological types were assigned on the basis of two qualifiers:
early-type\,/\,late-type\,/\,irregular, and disk\,/\,no
disk\,/\,irregular.  The division between late-type and early-type is
determined by the presence of spiral arms and the smoothness of the
image. The disk qualifier is based on the presence of a disk.  Because
faint, face-on disks in S0 galaxies are very difficult to detect,
early-type galaxies that were classified as diskless may contain disks
at low inclinations (see, e.g., Rix \& White 1990, J\o{}rgensen \&
Franx 1994).

If at least three out of four authors agreed on the value of the
qualifier the value was set to the median of the values of the four
authors.  If there was no such agreement, the type qualifier and / or
disk qualifier were set to ``uncertain''. Our classifications fit
directly into Dressler's (1980) scheme. Ellipticals are early-type
galaxies without a disk, S0s are early-type galaxies with a disk, and
spirals are late-type galaxies with a disk.  Galaxies with type
and\,/\,or disk qualifier ``irregular'' are classified as irregulars.
The classifications are listed in Table 1, and labeled in
Fig.~\ref{allgal.plot} [Plate 3].  Color representations of several
examples of galaxies of different morphologies are shown in
Fig.~\ref{selgal.plot} [Plate 4].



\setcounter{figure}{4}

\vbox{
\begin{center}
\leavevmode
\hbox{%
\epsfxsize=8cm
\epsffile{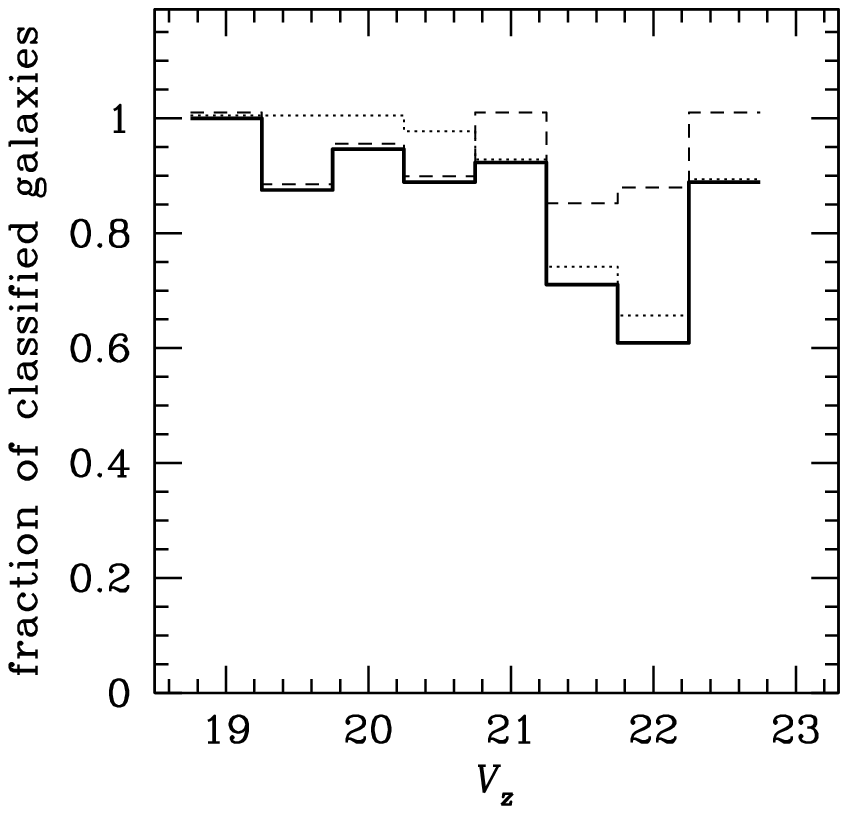}}
\begin{small}
\figcaption{
The agreement between visual
classifications of different authors.  The histograms show the fraction of
galaxies within a magnitude bin that were given identical
classifications by at least three out of four authors. 
The agreement on the type qualifier is indicated by the dashed line,
and the agreement on the disk qualifier by the dotted
line. The drawn line indicates the agreement on the type qualifier
and the disk qualifier simultaneously. The magnitude
bins are 0.5 magnitudes wide. Adjacent points are independent.
\label{class.plot}}
\end{small}
\end{center}}

The reliability of visual classifications has been discussed
extensively (e.g., Abraham et al. 1996a, Smail et al. 1997). It
appears that the robustness of the classification of HST images of
galaxies at $z \lesssim 0.5$ is not very different from that of the
classification of galaxies in nearby clusters observed from the
ground; even at low redshift it is difficult to distinguish between
adjacent Hubble types.  The robustness of the classification can be
assessed by comparing the classifications of different
authors. Figure~\ref{class.plot} illustrates the consistency of the
classifications.  Three of the four authors agree on $>85$\,\% of the
galaxies to $V_z = 21$.  The improvement in the consistency of the
classifications at magnitudes fainter than $V_z = 22$ is probably
real. The galaxy sample is dominated by irregular galaxies below this
magnitude, and these galaxies are more easily distinguished from the
other types than, e.g., ellipticals from S0s.



Some of the galaxies in the sample have faint, smooth spiral structure
(e.g., galaxy 328). In the adopted classification scheme, these
galaxies could either be classified as spirals (on the basis of the
spiral features), or S0s (on the basis of the absence of star
formation regions and dust).  It is not clear how these galaxies fit
into the Hubble sequence.  We will discuss the properties of these
galaxies in more detail in future papers on the morphology-density
relation, and the ``E+A'' galaxies in CL\,1358+62.  We note here that
four of these ambiguous galaxies were classified as S0s, the remainder
as spirals.

We compared the limiting magnitude of our classifications to the limits
adopted by Smail et al.\ (1997) and Ellis et al.\ (1997). Smail et al.\
classify galaxies to a limit (in signal-to-noise) $\sim 1$ magnitudes fainter
than in our study, and Ellis et al.\ to a limit $\sim 0.5$ magnitudes
fainter.

\subsubsection{Quantitative
Classification from Structural Parameters}

The visual classifications are by their nature not very quantitative,
nor can we assess systematic errors.  Because the vast majority of
cluster galaxies in the present sample are early-type galaxies, it is
desirable to have a quantitative indicator of the bulge-to-disk ratios
of these galaxies, to have an alternative way to discriminate between
bulge dominated and disk dominated galaxies.  As noted by Saglia et
al. (1997), the S\'ersic (1968) $n$ parameter provides a rough measure
of the bulge-to-disk ratio. The value of $n$ that gives the lowest
$\chi^2$ in a fit of the galaxy profile to an $r^{1/n}$ law depends on
the relative contributions of the bulge and the disk. If the disk
dominates, $n$ will be close to 1 (i.e. an exponential profile), if
the bulge dominates, $n$ will be close to 4 (i.e. the de Vaucouleurs
$r^{1/4}$ law).

All images of the cluster galaxies were fitted with two-dimensional
$r^{1/n}$-law models (S\'ersic 1968), with $1 \leq n \leq 4$, which
were convolved with the HST Point Spread Function (PSF). The PSFs were
generated with Tiny Tim 4.0b (Krist 1995). For each galaxy, a separate
PSF was used, appropriate to the galaxy position on the WF chip.  The
fitting method uses the full 2D information rather than an azimuthally
averaged light profile.  The fitting program determines the position,
ellipticity, position angle, effective radius $r_{\rm e}$, and the
surface brightness at the effective radius $\mu_{\rm e}$.  The method
is described for the $n=4$ case (the de Vaucouleur $r^{1/4}$-law) in
van Dokkum \& Franx (1996).  Models with $n=1,2,3,4$ were
fitted to the galaxy images.
The $\chi^2$ values of the four fits were compared to choose
the `best fitting profile type' for each galaxy.

We have performed simulations to test how well the $n$ parameter
correlates with bulge-to-disk ratio. Artifical galaxies were created
with varying bulge-to-disk ratios, inclinations, and effective radii
of the bulges and the disks. The effective radii of the disks span a
range from half the effective radius of the bulge to twice the
effective radius of the bulge.  In the simulations, the surface
brightness profiles of the bulges follow $r^{1/4}$ laws, and the
surface brightness profiles of the disks are exponential.  The
artificial galaxies were convolved with Tiny Tim PSFs, before
determining the $n$ parameter.  We used the same procedure to
determine the best fitting profile type $n$ as for the cluster
galaxies.

The simulations show that the $n$ parameter does indeed correlate with
bulge-to-disk ratio, but that it is also somewhat dependent on the
effective radius of the disk relative to the bulge: large, faint disks
relative to the bulge give rise to higher $n$ values than small,
bright disks.  Bulge fractions $\leq 20$\,\% give rise to $n \leq 2$
in $90$\,\% of the cases, and bulge fractions $\geq 70$\,\% always
give rise to $n \geq 3$ in our simulations.

In reality, galaxy profiles cannot always be described by the
combination of a de Vaucouleurs bulge and an exponential disk. As an
example, galaxy 86 is one of 2 galaxies that are best fitted by an
exponential profile ($n=1$), but are visually classified as
elliptical.  A closer examination of its surface brightness profile
reveals that it is best fitted by a combination of two exponentials.
Also, bars and other non-axisymmetric features complicate the
interpretation of the $n$ parameter.

In summary, the $n$ parameter is a rough measure of the bulge-to-disk
ratio, but one should be cautious in drawing conclusions for
individual galaxies on the basis of the $n$ parameter. We stress that
the $n$ parameter is not intended as a refinement of the visual
classifications, but as an independent and quantitative measure of the
morphology. We will express our results in the context of both the
visual classifications, and the best fitting profile type.

\subsubsection{Colors and Magnitudes}

The colors were measured from aperture photometry using the {\sc phot}
task in {\sc iraf}. For each galaxy, the radius of the aperture $r_{\rm c}$
was set equal to the effective radius, $r_{\rm e}$, as determined from the
fit to an $r^{1/4}$-law. The advantage of using the effective radius
rather than a larger aperture is that sky subtraction errors are
small.  Furthermore, this procedure minimizes the effects of color
gradients.

For many galaxies, the effective radii are $\lesssim 0\farcs 5$.
Within such small radii, the galaxy profiles are significantly
affected by the Point Spread Function (PSF).  The shape of the HST PSF
depends on the passband, and the measured colors within an effective
radius are affected by the differences between the $F606W$ and $F814W$
PSFs.  We determined the importance of this effect by constructing a
model galaxy with an effective radius of $0\farcs 3$, and convolving
it with $F606W$ and $F814W$ PSFs generated by Tiny Tim. The measured
color within $r_{\rm c}$ turned out to be $0.016$ magnitudes bluer
than the true color.  After deconvolution of the model galaxy, the
true color could be obtained within $0.001$ magnitudes. We tested whether
the Tiny Tim PSF is a reasonable approximation of the true PSF by
constructing a model galaxy using the $F606W$ and $F814W$ images of a star
in the field, and deconvolving it using a Tiny Tim PSF. Again, the true
color could be obtained within $0.001$ magnitudes after deconvolution.

To correct for the effects of the PSF,
we deconvolved all images before measuring
the colors. For the deconvolutions the {\sc clean} algorithm (H\"ogbom
1974) was used, which ensures flux conservation. For each galaxy
a PSF was created with Tiny Tim, appropriate for the position of
the galaxy on the chip. The deconvolution
does not influence the scatter in the CM relation significantly.
However, the deconvolution has a small, but systematic, effect on the
slope of the CM relation.

We performed extensive tests to establish how our conclusions depend
on the choice of aperture. We explored the effect of apertures with
radii $r_{\rm e}/2$, $2r_{\rm e}$, and $3r_{\rm e}$, as well as
apertures with a fixed angular radius ($0\farcs 2$, $0\farcs6$,
$1\farcs0$, and $1\farcs5$).  The systematic trends discussed in
Sect.~\ref{result.sec} also apply for all these apertures.  However,
the colors of individual galaxies may depend substantially on the
aperture size because some galaxies have significant color gradients.
We will return to this issue later.

``Total'' magnitudes were measured through $1\farcs 5$ radius
apertures.  The error in the total magnitude contributes to the
scatter in the CM relation.  This can be estimated at $|\alpha| \delta
V$, where $\alpha$ is the slope of the CM relation and $\delta V$ is
the error in the magnitude. Since $|\alpha| \sim 0.018$ (cf.\
Sect.\ \ref{cmeq.sec}) this contribution to the scatter in the CM
relation is negligible, for $\delta V < 1$ magnitude. The colors and
magnitudes of all 194 galaxies in the sample are listed in Table 1.

\subsubsection{Photometric Accuracy}

It is difficult to calibrate HST WFPC2 observations to higher absolute
accuracy than $5$\,\% (see, e.g., Holtzman et al. 1995, Whitmore
1997).  One of the main uncertainties is the charge transfer
efficiency (CTE) which leads to efficiency variations of a few percent
over individual WFPC2 chips (Whitmore 1997).  Fortunately, all
uncertainties that are effectively efficiency variations over the
WFPC2 field (such as the CTE problem) cancel out when determining
colors, provided the effect is not very wavelength
dependent. Therefore, in principle, color differences between galaxies
can be measured to very high accuracy despite limitations in the
absolute calibration.

The signal to noise (S/N) ratio in the HST observations of the
spectroscopically confirmed cluster members is very high; the formal
errors in the colors measured within $r_{\rm c}$ are $\lesssim
0.005$ magnitudes for all galaxies.  However, other systematic sources
of error remain, e.g., the flatfielding, the sky subtraction, and
uncertainties in the PSF.

The HST pointings were chosen to allow an overlap of $\sim 5''$
between adjacent observations. Therefore, a number of objects in the
CL\,1358+62 field were observed twice.  These repeated observations
allow us to directly assess the errors in the photometry.  We found 23
objects that were well exposed in two pointings; 15 of these are
spectroscopically confirmed cluster members.  The galaxy colors were
measured within $0\farcs7$ radius apertures, which is the median
$r_{\rm e}$ of the 194 cluster members.

The $1\,\sigma$ spread in the differences between the two color
measurements of the 23 objects is 0.015 magnitudes, implying an
uncertainty for a single observation of $0.015 / \sqrt{2} = 0.011$
magnitudes. The 23 objects span a similar range in magnitude as the
sample of 194 cluster members, and have a median magnitude of $V_z =
21$. We divided the sample into two bins to determine whether the
measurement uncertainty depends on the magnitude. For galaxies with
$V_z < 21$ the uncertainty is 0.009.  For galaxies with $V_z \geq
21$ the uncertainty rises to $0.017$ magnitude.

We noticed that the differences between the two observations were
often close to $-0.010$ or $+0.010$. This is almost certainly due to
systematic differences between the WFPC2 chips: after adding $0.010$
to the $F606W$ zeropoint of chip 3, the measurement uncertainty for
galaxies in the bright half of the sample was reduced to $0.002$
magnitudes.

The errors in the sky subtraction and the flat fielding are probably
largest near the edges of the chips. In addition, the shape of the PSF
has a strong positional dependence near the chip edges. Therefore, the
color measurement errors for galaxies near the centers of the chips
are likely to be smaller than the preceding estimates. The measured
scatter in the CM relation of the ellipticals and S0s ranges from
$0.021$ to $0.043$ (cf.\ Sect.\ \ref{cm.sec}), which is significantly
larger than the measurement error (0.011). Therefore, the contribution
of measurement errors to the measured scatter in the CM relation is
generally negligible. However, for consistency, we computed the
intrinsic scatter in the CM relation by removing an uncertainty of
$0.011$ magnitudes in quadrature from the measured scatter.

\section{The Color-Magnitude Relation}
\label{cm.sec}

\subsection{The Color-Magnitude Relation for Different Morphological
Types}
\label{cmeq.sec}

\begin{figure*}
\begin{center}
\leavevmode 
\hbox{%
\epsfxsize=\hsize
\epsffile[18 166 592 464]{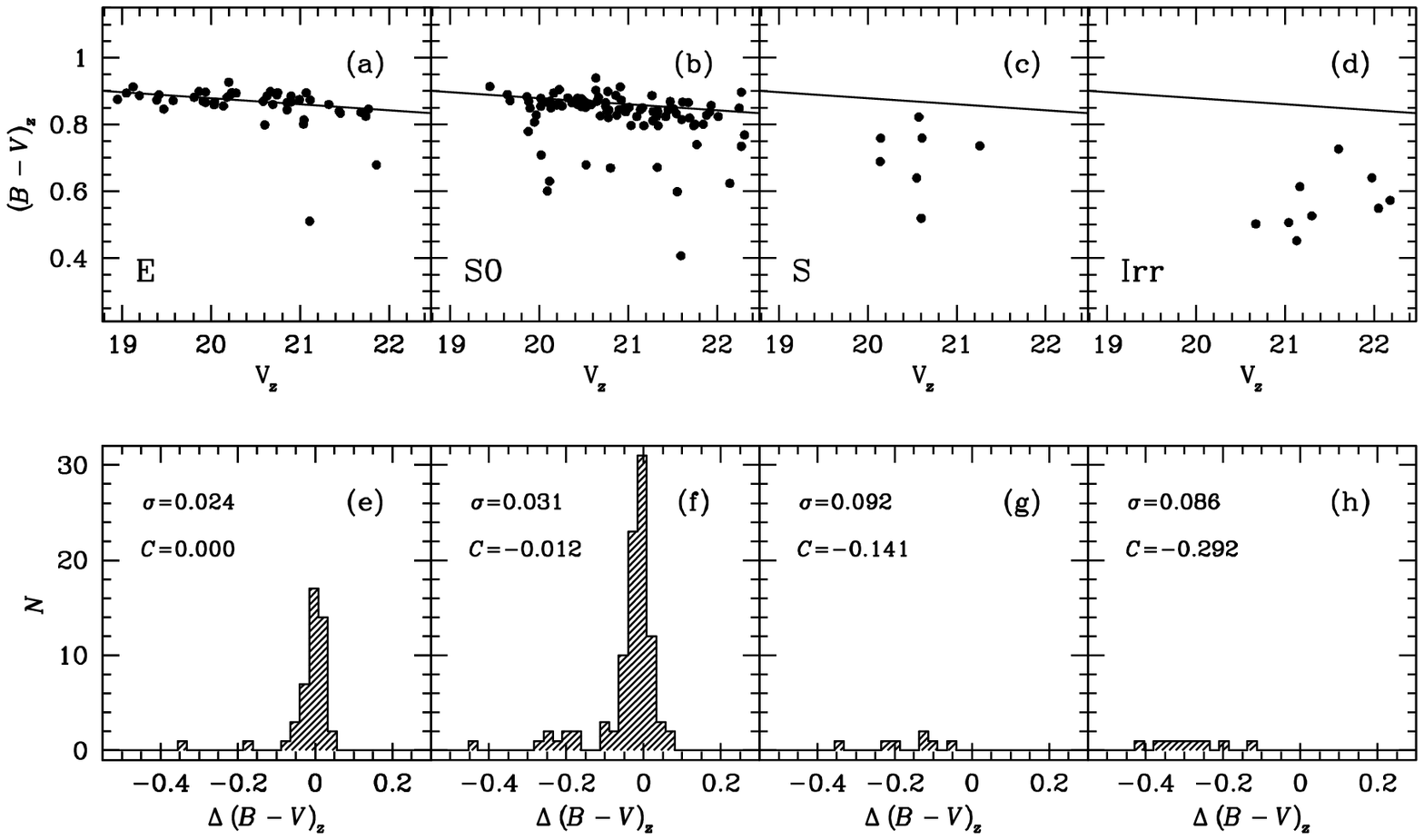}}
\begin{small}
\figcaption{\label{cm1.plot}The restframe $B-V$ color-magnitude relation
for galaxies of different morphological types (a-d). The line is a
least squares fit to the CM relation of the ellipticals, and is
repeated in each plot. Panels e-h show the distributions of the colors
after subtracting the CM relation of the ellipticals.  (see text).}
\end{small}
\end{center}  
\end{figure*}

The CM relations of cluster members of different morphological types
are presented in Figure~\ref{cm1.plot}(a-d).  The figure shows that
the CM relation depends strongly on morphological type.  Only the
ellipticals and S0s show a well defined relation, whereas the spirals
and the irregulars show a large scatter in the color.  Few ellipticals
are significant outliers from the ``ridge line'' of this relation, but
the fraction of outliers is higher for the S0s.  The distribution
about the ridge line appears to be tighter for the ellipticals than
for the S0s.


\begin{figure*}
\begin{center}
\leavevmode 
\hbox{%
\epsfxsize=\hsize
\epsffile[18 166 592 464]{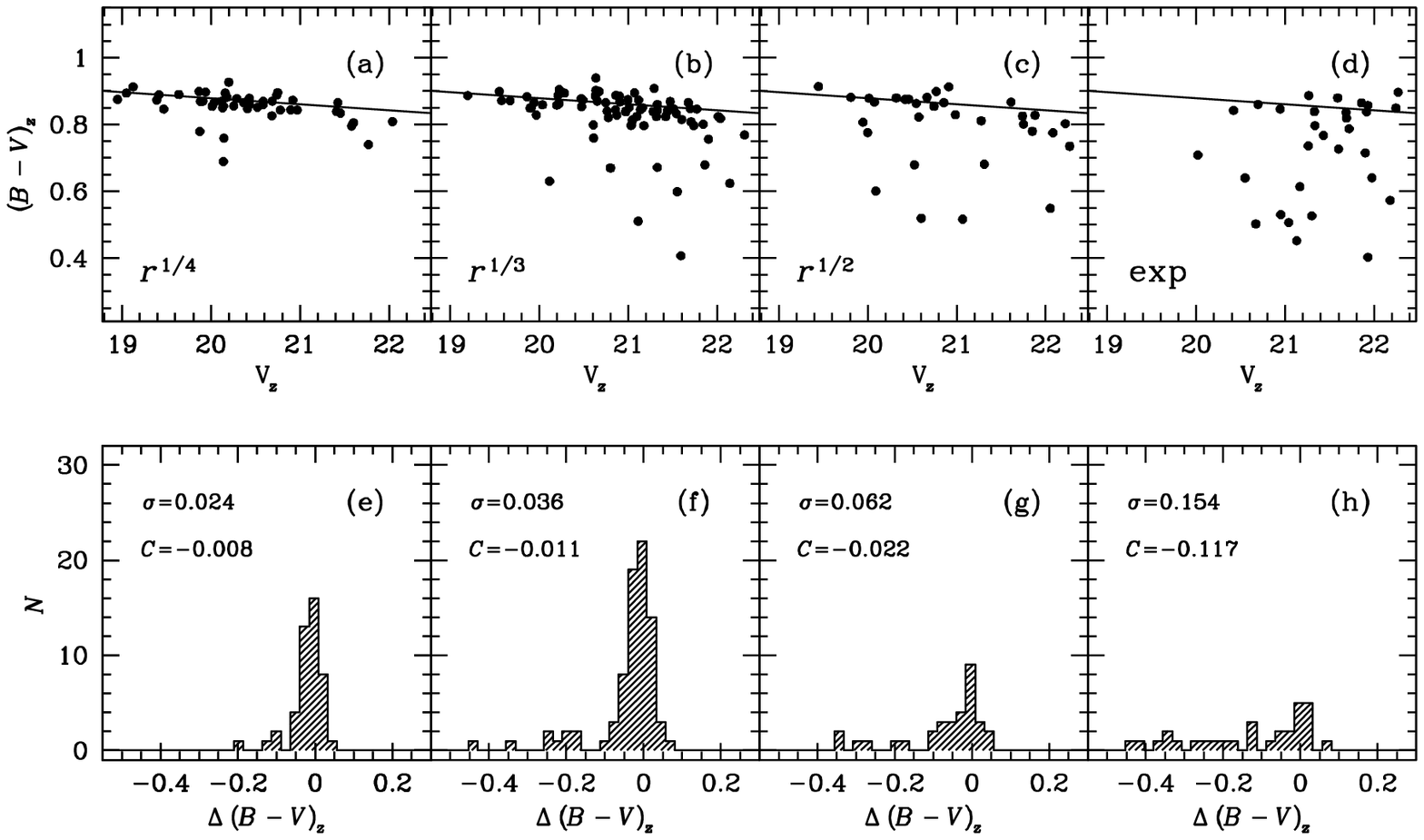}}
\begin{small}
\figcaption{\label{cm2.plot}The color-magnitude relation for galaxies
with different best fitting profile types. Galaxies are systematically
bluer and the scatter in the CM relation increases with decreasing
$n$.}
\end{small}
\end{center}  
\end{figure*}

\begin{figure*}[t]
\begin{center}
\leavevmode 
\hbox{%
\epsfxsize=\hsize
\epsffile[18 166 592 464]{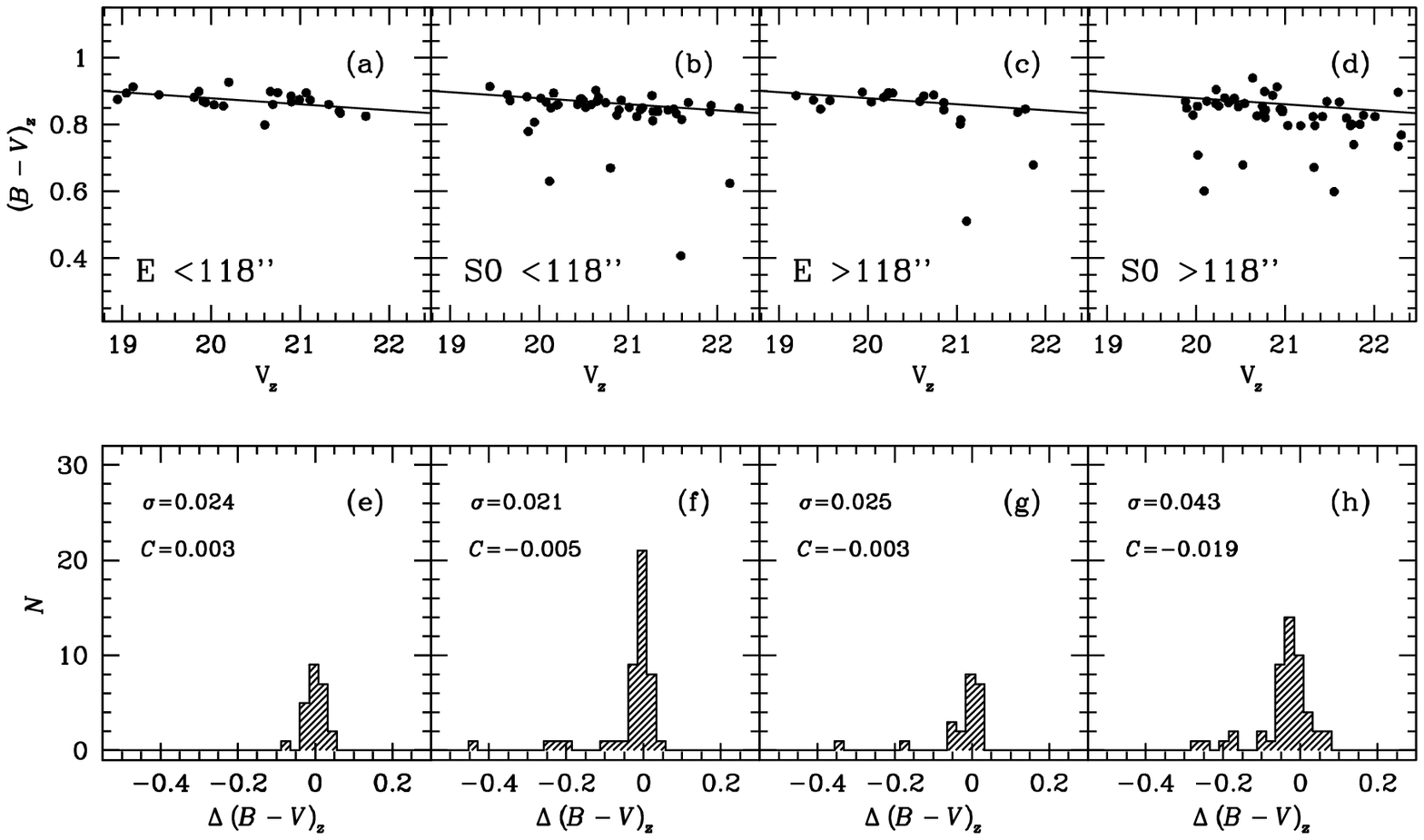}}
\begin{small}
\figcaption{\label{cm3.plot}Comparison of the color-magnitude relations
for ellipticals and S0s in the inner parts and in the outer parts of
the cluster.  The scatter in the CM relation of the S0s and the Es is
very similar inside $R=118''$.
Outside $R=118''$, the scatter in the S0 colors is larger
than that of the ellipticals by a factor 2.}
\end{small}
\end{center}  
\end{figure*}

\begin{figure*}[t]
\begin{center}
\leavevmode 
\hbox{%
\epsfxsize=\hsize
\epsffile[18 227 592 637]{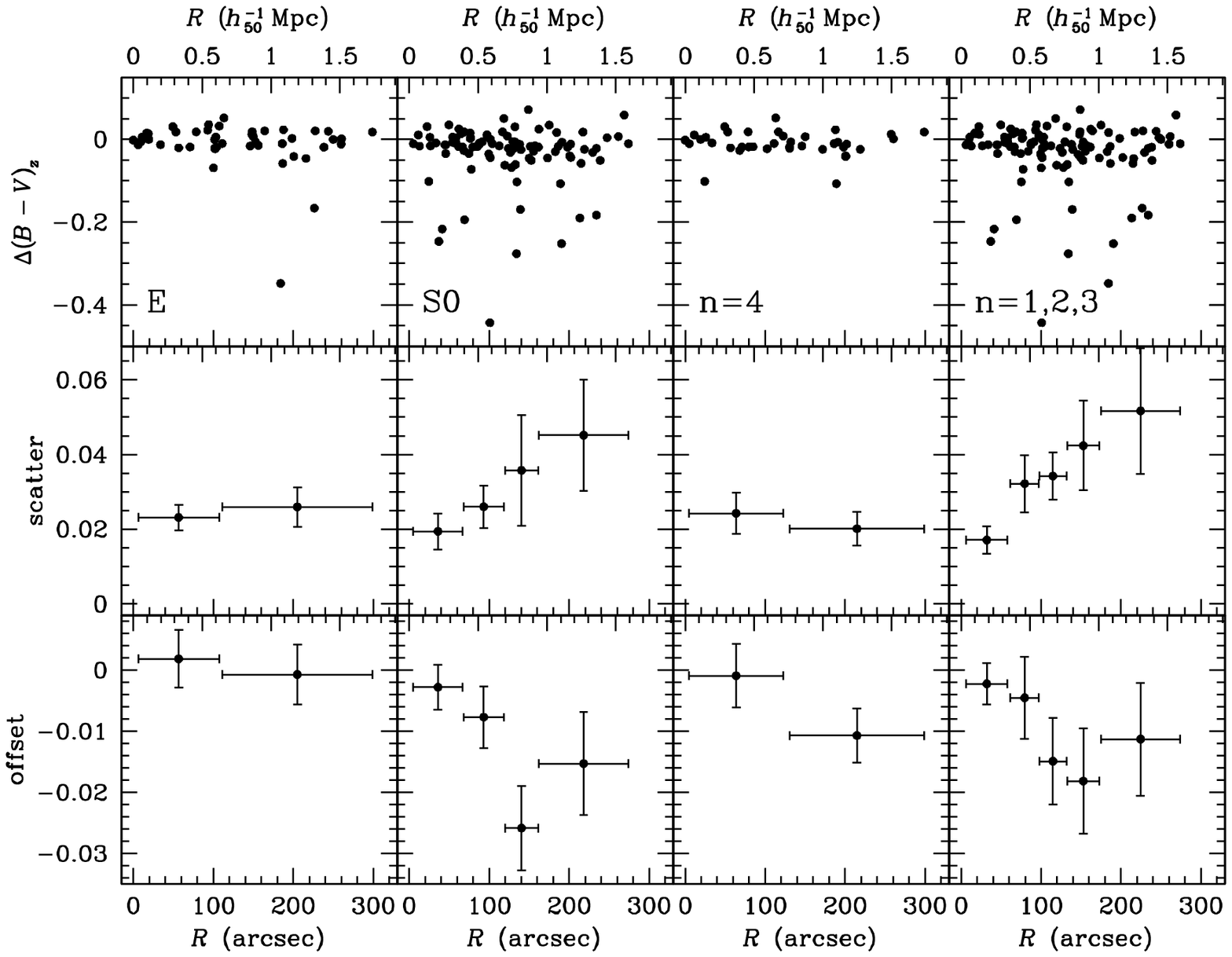}}
\begin{small}
\figcaption{\label{radcol.plot}The dependence of the residuals from the
CM relation on the distance from the BCG, for visually classified
ellipticals and S0s, and for early-type galaxies with $n=4$ and $n\leq 3$,
where $n$ is the value that gives the lowest $\chi^2$ in a fit to an
$r^{1/n}$ profile.  The scatter and offset determined with the
biweight estimators are calculated in radial bins, each containing 
at least 20 galaxies. Adjacent data points are independent.  The biweight
statistics give low weight to outliers so the scatter is not
a direct function of the number of outliers.  The scatter in the $(B-V)_z$
colors of the S0s and the early-types with $n\leq 3$ increases with
$R$. These galaxies are also bluer at larger distance
from the cluster center.  In contrast, the ellipticals and the
galaxies best fitted by an $r^{1/4}$-law do not show a significant
trend with radius, but the number of galaxies is too small to rule out
a change with radius in the scatter or mean color of $\lesssim 0.01$
magnitudes.}
\end{small}
\end{center}  
\end{figure*}

We quantified these effects as follows. First, we determined
the CM relation for the ellipticals with a least squares fit, excluding
the two bluest galaxies.  The form of this fit is
\begin{equation}
\label{cm.eq}
(B-V)_z = (0.866 \pm 0.004) - (0.018 \pm 0.005) (V_z-20.7).
\end{equation}
and it is shown in all upper panels of Fig. \ref{cm1.plot}.
For each galaxy we determined the residual color relative to the
fiducial CM relation of the ellipticals:
$\Delta (B-V)_z \equiv (B-V)_z + 0.018 V_z - 1.240$.
From the distributions of $\Delta (B-V)_z$ for the different
morphological types we determined the offset $C$ relative
to the CM relation of the ellipticals, and the scatter
$\sigma$. Histograms of the distributions of $\Delta (B-V)_z$
are shown in Fig.\ \ref{cm1.plot}(e-h).

The offset and the scatter were calculated using the biweight
estimators for the location and the scale of a distribution.  This
estimator was also used by Stanford et al.~(1997). The biweight
estimator gives higher weight to points that are closer to the center
of the distribution (see Beers, Flynn, \& Gebhardt 1990). Therefore,
the biweight estimator is insensitive to outliers.  For a Gaussian
distribution the biweight scale estimator reduces to the conventional
rms.  The values that we obtain are characteristic of
the scatter of the points close to the ridge line, and are insensitive
to the blue outliers.

Bower et al.~(1992b) and Ellis et al.~(1997) use the normalized median
absolute deviation (MAD) as a scatter estimator.  The biweight
estimator is more robust than the MAD estimator (Beers et al. 1990).
We checked if our results are sensitive to the choice of estimator by
computing the median and the normalized MAD for the distributions, and
comparing the results to the biweight estimators. In the present
study, the median and the normalized MAD give values very similar to
those obtained with the biweight estimators.

Table 1 and Fig. \ref{cm1.plot} show the offset and the scatter of the
CM relation. The error in the scatter was determined from bootstrap
resampling. The blue outliers were not excluded from the analysis,
although the biweight estimator gives them lower weight.
The visual impression of an increased scatter for the S0
galaxies when compared to ellipticals is confirmed
($0.031 \pm 0.004$ compared to $0.024 \pm 0.003$).
Furthermore, we find that the CM
relation of the S0 galaxies is offset to the blue with respect to that
of the ellipticals by $0.012 \pm 0.003$ magnitudes.  The significance
of the difference in the offsets of the ellipticals and the S0s can be
evaluated with the (non-parametric) Mann-Whitney test. The probability
that the ellipticals and the S0s are drawn from the same sample is
$<1$\,\%. The difference in the scatter for the ellipticals and S0s is
also significant, given the formal errors.  We directly
determined the significance of the difference between the
distributions of the residuals of the CM relations of the ellipticals
and the S0s with the Kolgomorov-Smirnov (KS) test.  The probability
that the distributions were drawn from the same parent sample
is $3$\,\%.

We tested whether the scatter and offset of the CM relation are
functions of the $n$ value derived from the surface brightness profile
fits.  In Fig.~\ref{cm2.plot}(a-d) the CM
relation is shown for different best fitting profile types, ranging
from $r^{1/4}$ ($n=4$) to exponential ($n=1$). The distributions of
the residuals from the CM relation are shown in
Fig.~\ref{cm2.plot}(e-h).  The offset and the scatter of the CM
relation vary systematically with $n$, towards a bluer offset and a
larger scatter for more disk dominated systems.  This trend is exactly
the same as the trend with morphology, confirming the distinction
between S0s and ellipticals.


\placefigure{cm2.plot}

This difference in properties between ellipticals and S0s in
intermediate redshift clusters has not been seen before. Similar
studies of the CM relation with HST have found very low scatter for both
ellipticals and S0s (e.g., Ellis et al.~1997). The main difference
with this study is that our large-area imaging extends beyond the
cluster core regions covered by previous studies with HST of intermediate
redshift clusters. This raises the question whether this effect is
related to distance from the cluster center.

\subsection{Radial Dependence of the Color-Magnitude Relation}
\label{result.sec}

We investigated whether the scatter in the CM relation of the
ellipticals and S0s depends on $R$, the distance from the
BCG.  As a first test, the galaxy sample was divided into two radial
bins.  The bin size, $118$ arcsec ($\sim 0.7$\,\h50min\,Mpc),
was chosen such that half of the
galaxies classified as early-type are contained in each bin.

Figure~\ref{cm3.plot} shows the CM relation of the ellipticals and S0s
inside and outside $R=118''$.  The scatter in the CM relation of the
ellipticals and the S0s in the inner regions of the cluster is very
similar: the observed scatter is $0.024 \pm 0.004$ for the
ellipticals, compared to $0.021 \pm 0.004$ for the S0s, and
the difference in mean color between the ellipticals and the S0s
within $R=118''$ is not significant.


Although the CM relations of the ellipticals and the S0s appear to be
similar in the core of the cluster, they are very different in the
outer parts of the cluster.  The CM relation of the elliptical
galaxies remains essentially unchanged, but the scatter for the S0s
increases substantially in the outer parts.  The CM relation of the
S0s in the outer parts is offset to the blue by $0.019$ magnitudes,
and the scatter is $0.043 \pm 0.009$, almost a factor 2 higher than
for the ellipticals in the outer parts, and the ellipticals and S0s in
the core.  The difference in the mean color between the S0s in the
inner parts and the S0s in the outer parts is significant at the
$95$\,\% confidence level.

The dependence of the CM relation on the distance to the cluster
center can be studied in more detail. In Fig.~\ref{radcol.plot} the
residuals of the CM relation are plotted against $R$, for visually
classified ellipticals and S0s, and for early-type galaxies separated
by best fitting profile type.  We experimented with the size of the
radial bins, and found that the results are robust when there are at
least $\sim 20$ galaxies in each radial bin.



There is a very clear trend of the scatter and mean color for the S0
galaxies, and for the early-type galaxies with $n\leq 3$. The trend is
much weaker, or absent, for the ellipticals and the early-type
galaxies best fitted by a de Vaucouleur $r^{1/4}$-law.
The scatter in the S0 colors increases from $0.017$ at $R = 40''$ to
$0.043$ at $R = 200''$. The intrinsic scatter increases from
0.013 to 0.042. The S0s in the outer parts of the cluster are $0.015$
magnitudes bluer than the S0s in the inner parts of the cluster.  The
scatter in the colors of the ellipticals is nearly constant with
radius, at $0.024$. The ellipticals in the outer parts of the cluster
do not appear to be much bluer than the ellipticals in the inner
parts.  However, the number of elliptical galaxies in the sample is
small. We cannot rule out a change in color of $\sim 0.01$ magnitudes.

These results are confirmed with the objective classification
provided by the best fitting profile types; the radial trend for the
early-types with $n\leq 3$ is very similar to the trend for the visually
classified S0s.  In contrast, there is no evidence for a radial trend
in the scatter for the galaxies best fitted by an $r^{1/4}$-law,
confirming the result for the visually classified ellipticals. The
early-types with $n=4$ in the outer parts are bluer than those in the
inner parts by $0.01$ magnitudes, but a Mann-Whitney test shows that
this difference is not statistically significant.

One might be concerned that the gradient in the S0 population is
driven by the radially increasing fraction of blue outliers.
We tested the effect of outliers by excluding from the analysis the S0
galaxies that are more than 0.1 magnitudes bluer than the CM relation.
We stress that since we have full membership information
there is no justification for removing the blue outliers from the sample,
other than to test whether these galaxies drive the trends with radius.
The trend in the S0 colors proved to be very similar for $R < 150''$.
The scatter in the outermost bin goes down from $0.046$
to $0.031$ when the four blue outliers in this bin are removed,
indicating that the scatter in individual radial bins has considerable
uncertainty.
The robustness of the trends in the scatter and the mean
color demonstrate that the scatter and mean color are a good
approximation to the width and location of the CM ridge line.  Our
results therefore describe the bulk of the population, and are not
unduely influenced by a small interfering population.

We tested whether the trends with radius of the S0 galaxies could be
caused by the misclassification of a few early-type spirals as S0s in the outer
part of the cluster. Assuming a scatter of $0.021$ for the S0 population,
and a three times larger scatter for Sa galaxies, we find that $\sim
55$\,\% of the S0 galaxies in the outer parts of the cluster must be
misclassified spirals to explain the observed scatter of $0.043$
magnitudes.
We conclude that our conclusions are robust against the misclassification
of a few spirals as S0s.

\vbox{
\begin{center}
\leavevmode
\hbox{%
\epsfxsize=8cm
\epsffile{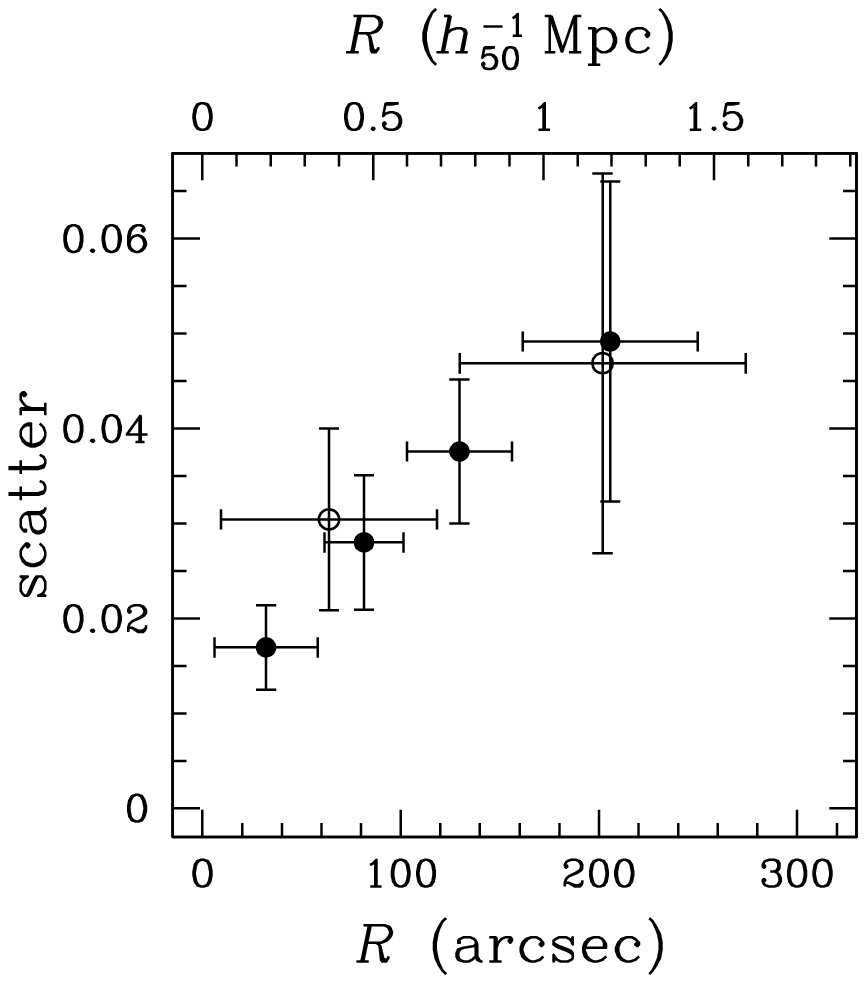}}
\begin{small}
\figcaption{\label{n34r.plot}The dependence of the scatter in the CM
relation on $R$, the distance from the BCG, for early-type galaxies
with $n=3$ (solid circles), and $n=1,2$ (open circles), with $n$ the
value that gives the lowest $\chi^2$ in profile fits of the form
$r^{1/n}$. The $n=3$ galaxies often have strong bulges. The trend with
$R$ of the $n=3$ galaxies is very similar to the trend of the $n\leq 2$
galaxies.  This shows that the trend of the colors of the S0 galaxies
with $R$ is not driven by bulge-to-disk ratio.}
\end{small}
\end{center}}

We also tested whether the gradient in color and scatter in the CM
relation is caused by a gradient in bulge-to-disk ratio: it may be that there
are more disk dominated galaxies in the outer parts of the cluster, and that
the disks are bluer than the bulges.
When we subdivide our sample by best fitting profile parameter $n$, we
find that the gradients are independent of $n$, for $n\leq 4$. In
particular, the trends with $R$ are very clear in the $n=3$ galaxies,
which have significant bulges. In Fig.~\ref{n34r.plot}, the radial
trend of the scatter in the CM relation is plotted for galaxies
classified as early-type with $n=3$ (solid circles), and $n\leq 2$ (open
circles). The colors of the S0 galaxies with strong bulges ($n=3$)
depend similarly on $R$ as the colors of S0s with weak bulges ($n\leq 2$).
This result strongly suggests that both bulges and disks are
responsible for the color differences. We return to this issue in
Sect. \ref{colgrad.sec}.

\begin{figure*}[t]
\begin{center}
\leavevmode 
\hbox{%
\epsfxsize=\hsize
\epsffile[30 519 547 677]{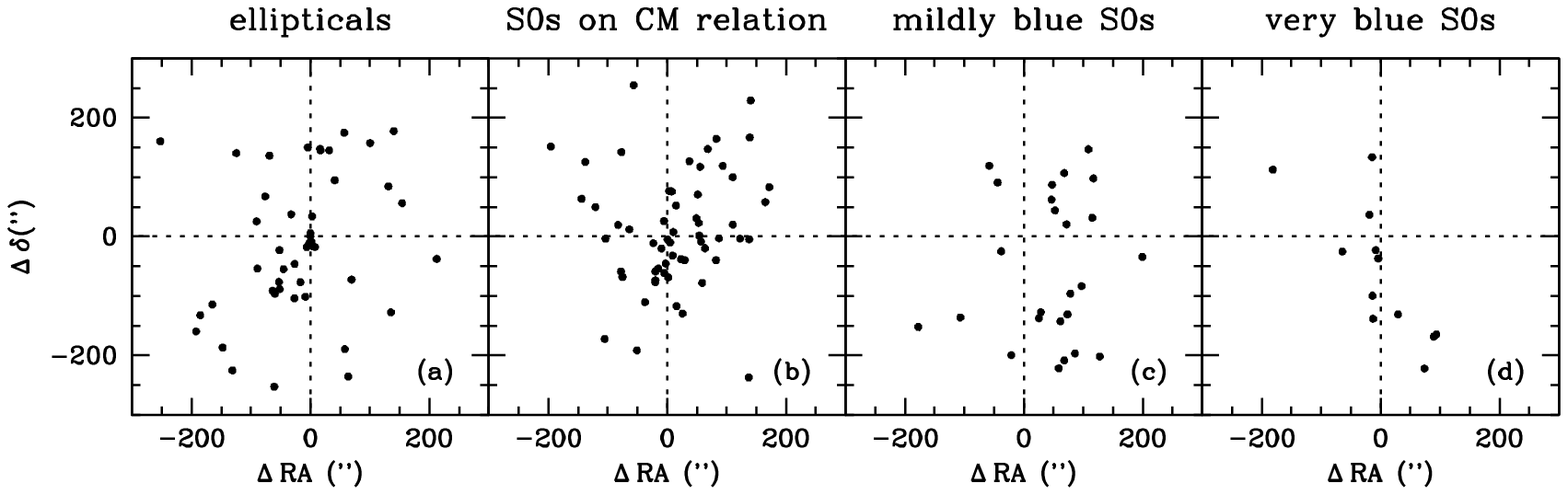}}
\begin{small}
\figcaption{\label{spatial.plot}The spatial distributions of the
ellipticals (a), and the S0s in various color bins (b-d).  In (b), the
distribution of S0s within 0.021 magnitudes (the $1\,\sigma$ scatter
in the CM relation of the S0s in the inner parts of the cluster) of
the CM relation is shown.  In (c), mildly blue ($-0.10 < \Delta (B-V)_z <
-0.021$ are shown, where $\Delta (B-V)_z$ is the distance to the CM
relation) S0s are shown, and (d) shows
the distribution of very blue S0s ($\Delta (B-V)_z \leq
-0.10$). The distributions of the various subsamples are very
different. In contrast to the S0s on the CM relation,
the mildly blue S0s avoid the cluster core.}
\end{small}
\end{center}  
\end{figure*}

We investigated whether either the red or blue galaxies are located in
subclumps. Figure \ref{spatial.plot} shows the spatial distributions
of the S0s in various color bins. For comparison, the spatial
distribution of the ellipticals is also shown.  The differences
between the spatial distributions of the various subsamples are
striking. The blue S0s avoid the cluster core, and their spatial
distribution does not appear relaxed. There seem to be more
mildly blue S0s to the West of the cluster center than to the East.
These results confirm that the S0 galaxies in the core are redder than
the S0 galaxies in the outskirts of the cluster. 



We conclude that the differences between the CM relations of the
ellipticals and the S0 galaxies discussed in the previous Section are
caused by the increased scatter and bluer colors of the S0s in the
outer part of the cluster.  The S0s in the inner parts follow a
similar CM relation as the ellipticals, whereas those in the outer
parts have characteristics indicative of more recent, or more intense,
star formation.  We cannot exclude a small ($\leq 0.01$) trend of the
elliptical colors with radius, but the trend of the S0 colors is much
more significant.  These are the key observational results of the
paper; they are discussed more extensively below.

\section{Blue Bulges and Disks}
\label{colgrad.sec}

It is natural to expect that the color differences between S0 galaxies
are caused in a large part by blue disks, as it is usually assumed
that star formation in disks lasts much longer than star formation in
bulges.  Some bulge dominated galaxies also show blue
colors, hence the process must be more complex.  The high resolution of our
HST data allows us to study the radial color gradients in the S0
galaxies, and to directly test if the central parts behave differently
than the outer parts.

As a first test, we determined the colors within 0.5 $r_{\rm e}$,
instead of the 1 $r_{\rm e}$ used above.  The color changes going to a
smaller aperture do not appear correlated with radius in the
cluster.  We then examined the full color profiles of our
sample galaxies. In general, the color gradients within the galaxies are
small, but we find some galaxies with disks that are considerably bluer
than their bulges or the reverse.

We show three typical color profiles in
Fig.~\ref{colgrad.plot}\,(c). Color images of these galaxies are shown
in Fig.\ \ref{selgal.plot} [Plate 4].  The radial surface brightness
profiles of the galaxies are shown in Fig.~\ref{colgrad.plot}\,(a).
These profiles were determined from the deconvolved images with the
{\sc galphot} package (Franx, Illingworth, \& Heckman 1989).  The top
profile in Fig.~\ref{colgrad.plot}\,(c) is a normal S0 on the CM
relation, the middle profile is an S0 which is slightly blue, and the
bottom profile is a very blue S0.  Clearly, these color profiles do
not show large radial changes.



\vbox{
\begin{center}
\leavevmode
\hbox{%
\epsfxsize=8cm
\epsffile{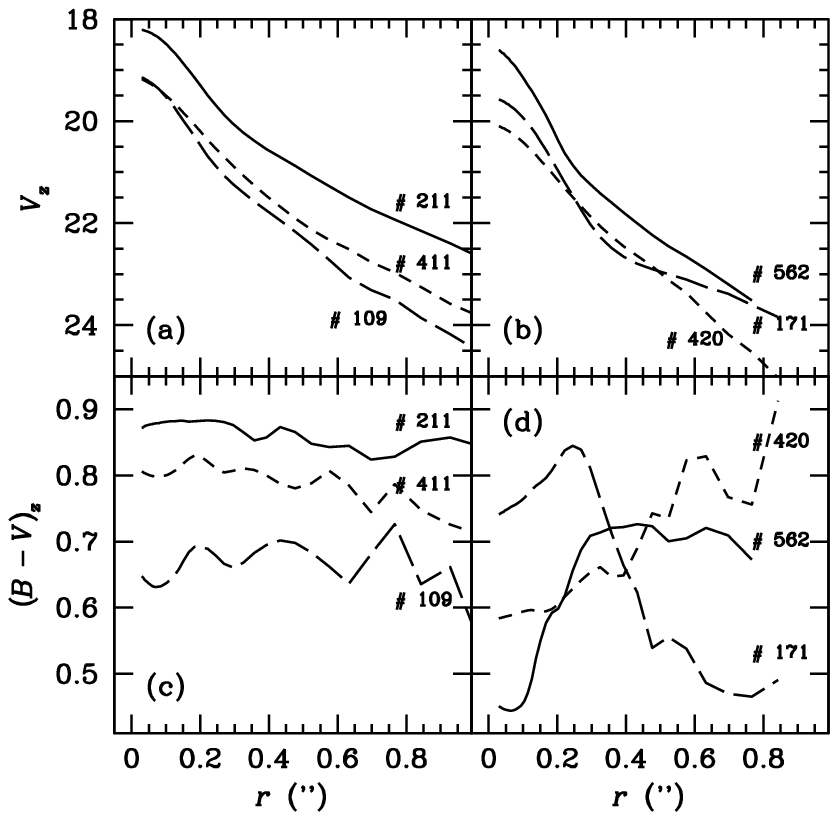}}
\begin{small}
\figcaption{\label{colgrad.plot}Radial surface brightness profiles
(a,b) and color profiles (c,d) of early-type galaxies in CL\,1358+62.
In (a) and (c), surface brightness profiles
and typical color profiles of an S0 galaxy on the
CM relation (211), a mildly blue S0 (411), and
a very blue S0 (109) are shown.
The color gradients of these S0
galaxies are very similar, indicating that both the bulges and
the disks of the blue S0s contain young populations. The majority
of the blue S0s have shallow color gradients. In (b) and (d), examples
of galaxies with extreme color gradients are shown: two galaxies with blue
bulges compared to their disks (420 and 562), and a galaxy with a blue disk
compared to its bulge (171). Images of the galaxies in this Figure are
shown in Fig. \ref{selgal.plot} [Plate 4].}
\end{small}
\end{center}}

Although the color differences between the bulges and the disks are
small for most galaxies, some galaxies do have large color
gradients. In particular, some of the bluest early-type galaxies have
very blue bulges. In Fig.~\ref{colgrad.plot}\,(d) three examples of
galaxies with extreme color gradients are shown.  One has a blue disk,
two have blue bulges.  Images of the galaxies are shown in
Fig.~\ref{selgal.plot} [Plate 4].  At all radii these galaxies are
bluer than galaxy 211, which falls on the CM relation.

We conclude that the trends in the S0 colors with distance from the
BCG cannot solely be attributed to prolonged star formation in the
disks of the S0s in the outer part of the cluster.  Although there are
a few galaxies with blue disks and red bulges, most S0 galaxies have
small color gradients, implying that the bulges were affected as much
as the disks, or that the star formation was stronger in the centers
of the disks. Interestingly, we find that some of the bluest galaxies
have very blue bulges compared to their disks.

\section{Are There Luminous Blue Galaxies ?}
\label{bluegal.sec}

Fig.~\ref{cm1.plot} demonstrates that the brightest galaxies in
CL\,1358+62 are all red. There are no galaxies that are more than 0.1
magnitudes bluer than the CM relation among the brightest 29 galaxies
in the sample.  Their low scatter in color suggests that the
population of bright galaxies is very stable between $z=0.33$ and
$z=0$.  The relatively low luminosity of the bluest galaxies has been
noticed before in other clusters (e.g., Butcher \& Oemler 1984,
Thompson 1986).

Interestingly, the distribution along the ridge line of the CM
relation of the bright S0 galaxies also seems tighter than that of the
less luminous S0s.  Also, judging from Fig.~\ref{cm2.plot} it appears
that the radial behavior of the colors of the brightest S0s is more
similar to that of the ellipticals than to that of the fainter S0s,
although the numbers are small.

We measured the scatter and offset of the CM relation of the brightest
one third of the S0 galaxies ($V_z < 20.52$).  The scatter in this
sample is $0.019 \pm 0.003$, and the offset $-0.003 \pm 0.004$.  These
values are similar to the scatter and offsets of the ellipticals
(cf. Table 2) and of the S0s in the inner part of the cluster. The
scatter for the bright S0s is significantly different from that of the
whole S0 sample.  Furthermore, the S0s with $V_z < 20.52$ do not show
the strong trend with radius that is evident for the whole S0
sample. The scatter for the bright S0s at $R>118''$ is $0.021 \pm
0.004$, and the offset $-0.010 \pm 0.005$.  The scatter for the
complete S0 sample in the outer parts is much higher, at $0.043 \pm
0.009$.

Although the brightest galaxies are generally very red, there are a
few luminous early-types that are slightly bluer than the average CM
relation.  As an example, galaxy 233 is the 8th brightest galaxy in
the sample, and is 0.04 magnitudes bluer than the CM relation. It is
classified as an elliptical, and it is best fitted by a de Vaucouleur
$r^{1/4}$-law.  (see Fig.~\ref{selgal.plot} [Plate 4]).  It may be
that the brightest galaxies in the cluster have had a different star
formation history than the bulk of the cluster population, e.g., they
may have had their star formation turned off at earlier times.  The
presence of a few bright galaxies with slightly bluer colors suggests
that we may find luminous blue early-types in
clusters at higher redshifts.

\section{Implications for the Star Formation Histories}

In this Section, we construct a range of models for the star formation
histories of the ellipticals and the S0s. The aim is to constrain the
most recent period of star formation in the ellipticals and the S0s
from the observed scatter in the CM relation.  We assume that the
scatter in the CM relation arises from age differences among the
galaxies.  The fact that essentially all early-type galaxies that are
more than 0.1 magnitudes bluer than the CM relation have strong Balmer
absorption lines (Fisher et al.~1997) supports this interpretation.

\subsection{Models}

While the usual assumption is that most of the star formation in
early-type galaxies occurred in single bursts, there is good evidence
that many galaxies in clusters have more complicated star formation
histories.  The spectra of
blue galaxies in intermediate $z$ clusters
can be fit by models that incorporate truncated star formation in
disks, or secondary bursts of star formation in spirals or ellipticals
(e.g., Couch \& Sharples 1987, Poggianti \& Barbaro 1996, Barger et
al. 1996).

Motivated by these studies, we consider three scenarios: 1) formation
of all the stars in a galaxy in a single burst of star formation, 2)
formation of the bulk of the stars in a first burst, followed by a
second burst of star formation involving a small fraction of the mass
of the galaxy, and 3) a truncated uniform star formation rate.  We use
simple semi-analytical descriptions of the color evolution to predict
the scatter in the CM relation.  The derivation of the color evolution
is given in Appendix A for each of the models.

The purpose of the modeling is to compare the model predictions to the
observed scatter in the CM relation, under various assumptions. We
focus on the scatter in the CM relation of the ellipticals and of
the S0s in the outer parts of the cluster, since the scatter in the CM
relation of the S0s in the inner parts is very similar to that of the
ellipticals.  The intrinsic scatter for the ellipticals is 0.022
magnitudes, and for the S0s in the outer parts of the cluster 0.042
magnitudes (cf. Table 2).  A small ($10$\,\%) correction must be made
to the intrinsic scatter because the increased luminosity of young
galaxies artificially increases the scatter in the CM relation
(cf.~Appendix B). The corrected intrinsic scatter is $0.020$
magnitudes for the ellipticals, and $0.038$ for the S0s in the outer
parts of the cluster.

\subsection{Application of the Models to the Data}

\subsubsection{Constraints on the Most Recent Period of Star Formation}
\label{mod.sec}

The scatter in the CM relation only constrains the scatter in the
relative ages of the galaxies, and not the absolute ages (see Bower et
al. 1992b, Appendix A). The galaxies could have an arbitrarily low
mean age, provided that their formation was sufficiently
synchronized. It is not possible to determine the most recent period
of star formation from the models without making further assumptions.

To break this degeneracy, we assume simple top hat probability
distributions in time for the star formation events that characterize
the models (single bursts, secondary bursts, truncation of uniform
star formation). In Sect.\ \ref{difprob.sec} we will investigate the
effects of relaxing this assumption. In the single burst model,
galaxies form in a $\delta$-peak at a random time.  In the secondary
burst model, $80$\,\% of the mass of the galaxies forms in a burst at
$t=0$, and $20$\,\% in a second burst at a random time.  In the
truncated star formation model, the star formations starts at $t=0$
for all galaxies, continues at a constant rate, and terminates at a
random time.

\begin{figure*}
\begin{center}
\leavevmode 
\hbox{%
\epsfxsize=\hsize
\epsffile[18 144 592 718]{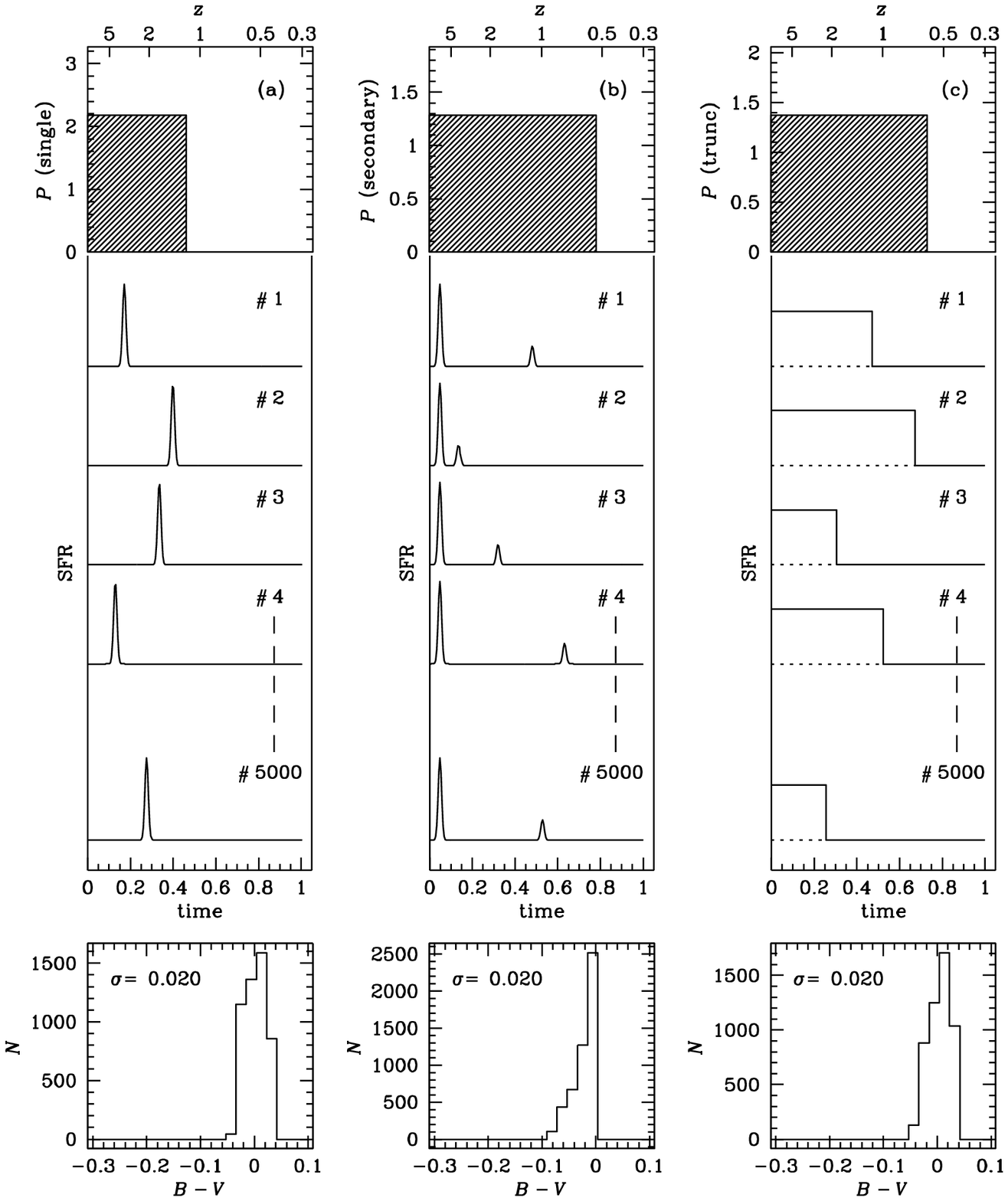}}
\begin{small}
\figcaption{\label{mod1.plot}
Reproduction of the scatter in the CM relation of the ellipticals (0.020) in
three different models. In (a), galaxies form in a $\delta$-peak at
a random time between $t=0$ and $t=0.46$ ($z=1.2$; $q_0 = 0.5$).
In (b), $80$\,\%
of the mass of the galaxies forms in a single burst at $t=0$, and
$20$\,\% of the mass in a second burst at a random time between
$t=0$ and $t=0.78$ ($z=0.5$). In (c), the galaxies start forming stars
at $t=0$ and stop forming stars at a random time between $t=0$ and
$t=0.73$ ($z=0.6$). Note that $t$ is scaled such that $t=1$ corresponds to
$z=0.33$. The top panels show the probability distributions $P$
for the events that characterize the models (single bursts, secondary bursts,
truncation of star formation).
The model color distributions at $z=0.33$
are shown in the bottom panels, derived from Monte Carlo simulations with
5000 model galaxies.
The constraints on the last period of star formation are very dependent
on the model for the star formation history.}
\end{small}
\end{center}  
\end{figure*}

\begin{figure*}
\begin{center}
\leavevmode 
\hbox{%
\epsfxsize=\hsize
\epsffile[18 144 592 718]{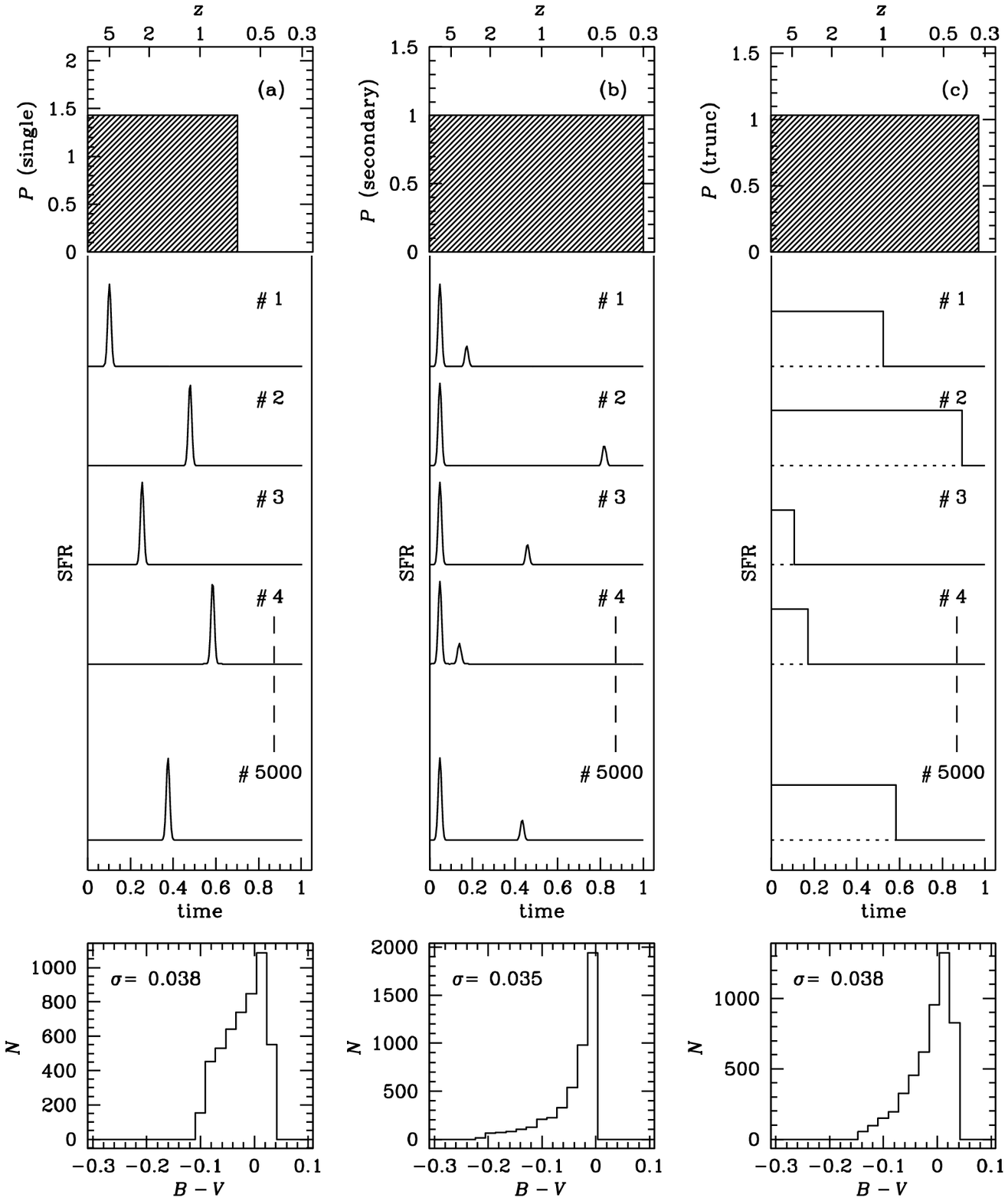}}
\begin{small}
\figcaption{\label{mod2.plot}
Reproduction of the scatter in the CM relation of the S0 galaxies
in the outer parts (0.038) in three different models. In the single burst
model, the youngest S0s formed at $z \approx 0.7$ (a). Secondary
bursts involving 20\,\% of the mass that occur from $t=0$ to the epoch of
observation (b) predict a scatter that is somewhat
lower than what is observed.
The truncated star formation model (c) can reproduce the observed
scatter if the youngest S0s were forming stars up to $z=0.33$,
the redshift of the cluster.}
\end{small}
\end{center}  
\end{figure*}

The models are schematically represented in Fig.~\ref{mod1.plot} and
Fig.~\ref{mod2.plot}.  The scatter in the CM relation of the
ellipticals is reproduced in Fig.~\ref{mod1.plot}, and that of the S0s
in the outer part of the cluster in Fig.~\ref{mod2.plot}.  The bottom
panels in Fig.~\ref{mod1.plot} and Fig.~\ref{mod2.plot} show the
histograms of the residuals from the CM relation, derived from
Monte-Carlo simulations with 5000 galaxies. The scatter and
offset of these simulated distributions
are measured in the same way as the observed parameters.


The scatter of 0.020 in the colors of the ellipticals can be
reproduced by the three models (Fig.~\ref{mod1.plot}).  In the single
burst model, the youngest ellipticals formed at $z = 1.2$. In the
secondary burst model, the most recent bursts occurred at $z=0.5$. In
the truncated star formation model, the youngest ellipticals were
forming stars up to $z=0.6$. The constraints on the most recent period
of star formation thus depend on the assumed model for the star
formation history. In particular, in the truncated star formation
model the ages of the most recently formed galaxies are a factor of 2
lower than in the single burst model.  This generic feature of the
models can be derived analytically (cf.~Appendix A).

The scatter in the colors of the S0s can be reproduced by the single
burst model and the truncated star formation model
(Fig.~\ref{mod2.plot}). In the single burst model, the youngest S0s
formed at $z=0.7$. The secondary burst model predicts a scatter of
0.035, if the mass fraction involved in the burst is $20$\,\% and the
bursts occur continuously up to $z=0.33$. This is slightly lower than
the observed scatter. Stronger bursts (involving mass fractions of
$\sim 25$\,\%) are required to explain the observed scatter with
this model. The truncated star formation model can reproduce the
observed scatter if the youngest S0s were forming stars up to
$z=0.33$, the redshift of the cluster.

An important result is that all three models allow for quite recent
star formation in the S0s in the outer part of the cluster. The
formation time of the galaxies is pushed back furthest in the single
burst model, but even in this model the youngest S0s formed as
recently as $z \sim 0.7$.

\subsubsection{Other Probability Distributions}
\label{difprob.sec}

The assumption in the preceding Section is that the probability
distributions for the star formation events that characterize the
models are simple top hats. Here, we investigate how stable our
conclusions are when this assumption is relaxed. We consider two other
probability distributions: an exponentially decreasing probability
distribution, and a Gaussian. For simplicity, only single burst models
are considered, and only the scatter in the ellipticals and the S0s in the
core of the cluster is reproduced.

In Fig.\ \ref{dist.plot} three probability distributions are shown
that reproduce the scatter in the ellipticals and S0s in the core of
CL\,1358+62. The exponential distribution has the form $P = e^{-5
t/t_{0.33}}$, where $t_{0.33}$ is the age of the universe at
$z=0.33$. The mean of the Gaussian is at $z=2$, and the width is $0.13
t_{0.33}$.

The formation time of the youngest galaxies is more difficult to
define for other distributions than the top hat, since the Gaussian
and exponential distribution have a tail extending to infinity. We
calculated when 50\,\% and 75\,\% of the galaxies have formed assuming
the three distributions (top hat, exponential, and an example of a
Gaussian).  The open dots indicate the time when 50\,\% of the
galaxies have been formed, and the solid dots indicate the time when
75\,\% of the galaxies have been formed.  In all the models $\sim
25$\,\% of the ellipticals and the S0 galaxies in the core of the
cluster formed after $z=2$. This is a generalization of the result
that the youngest ellipticals and S0 galaxies in the core of the
cluster formed at $z\sim 1.2$, which was derived for the top hat
model in the previous Section.



\vbox{
\begin{center}
\leavevmode
\hbox{%
\epsfxsize=8cm
\epsffile{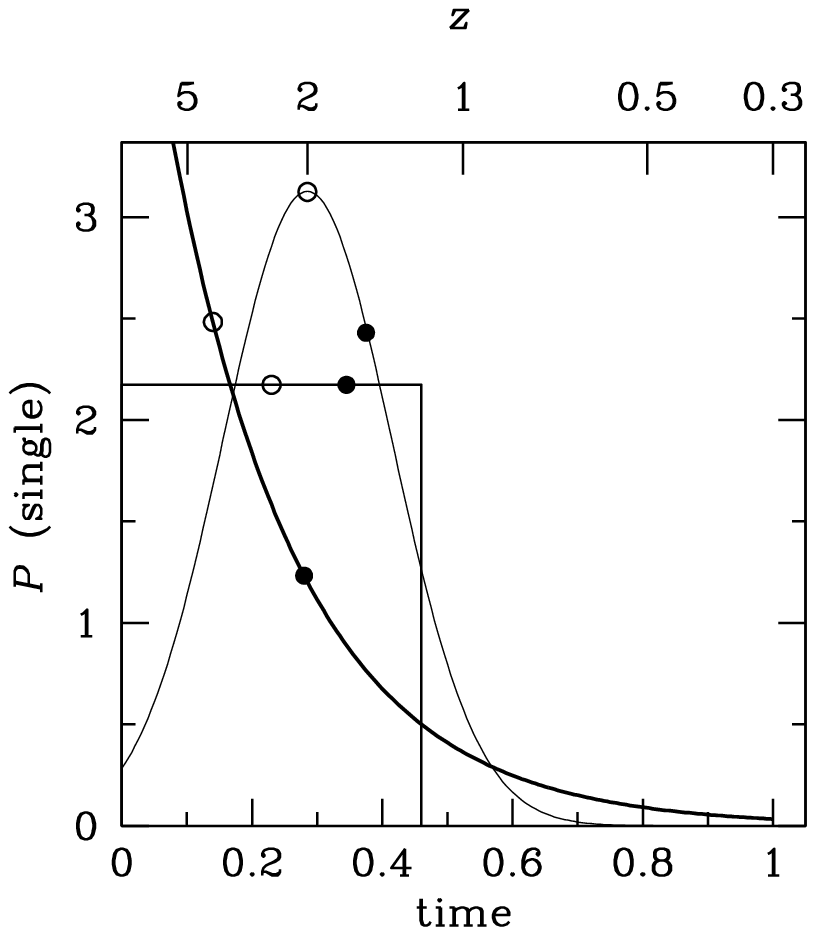}}
\begin{small}
\figcaption{\label{dist.plot}
Three probability distributions for galaxy formation in
single bursts of star formation that reproduce the observed scatter
in the CM relation of the S0 galaxies in the core and the elliptical
galaxies in CL\,1358+62. Open dots indicate the times when 50\,\% of
the galaxies have been formed, in each of the models. Solid dots indicate
the times when 75\,\% of the galaxies have been formed.}
\end{small}
\end{center}}

Similar to the top hat distribution, the exponentially decreasing
probability distribution has one free parameter, and is therefore
fully determined by the observed scatter at $z=0.33$. The Gaussian
distribution has two free parameters: any Gaussian distribution that
satisfies the constraint $\sigma_t / (t_{0.33} - t') = 0.18$, where
$\sigma_t$ is the width of the Gaussian and $t'$ is the mean,
reproduces the observed scatter in the CM relation for the ellipticals
and the S0s in the core of the cluster.  We note that the Gaussian
distribution has the general property that it allows for more recent
star formation than the top hat or the exponential.  In particular,
the time when $75$\,\% of the galaxies have been formed is always
after $z=2$, for any combination of $\sigma_t$ and $t'$.

\subsubsection{Constraints on the Scatter in the Luminosity Weighted Age}

The constraint on the most recent period of star formation is very
dependent on the model that is assumed for the star formation history
(cf.\ Sect.\ \ref{mod.sec}). In particular, models with truncated
continuous star formation allow for much later star formation than
single burst models.  Here, we investigate how well we can constrain
the scatter in the mean ages and the mean luminosity weighted ages of
the galaxies.

The scatter in the mean ages of the galaxies is also rather model
dependent. In the models described in Sect. \ref{mod.sec}, $0.05 <
\Delta \tau / \langle \tau \rangle < 0.20$ for the ellipticals, and
$0.07 < \Delta \tau / \langle \tau \rangle < 0.35$ for the S0s in the
outer parts of the cluster.  This large range results from the fact
that the color of a galaxy is mostly determined by the youngest stars
that contribute most of the light.

The scatter in the $V$ band luminosity weighted ages is better
constrained. The expressions for the luminosity weighted ages $\tau_L$
for each of the models are given in the Appendix.  For the
ellipticals, the single burst model gives $\Delta \ln \tau_L = 0.18$,
the truncated star formation model $\Delta \ln \tau_L = 0.16$, and the
secondary burst model $\Delta \ln \tau_L = 0.13$.  For the S0s in the
outer parts, we find $\Delta \ln \tau_L = 0.35$ in the single burst
model, and $\Delta \ln \tau_L = 0.30$ in the truncated star formation
model.

Secondary burst models provide the strongest constraints on the
scatter in the luminosity weighted ages; weaker but similar
constraints are provided by the single burst model and in the
truncated star formation models.  The constraints on $\Delta \ln
\tau_L$ are not very dependent on the distributions of the formation
times and truncation times. The various probability distributions
discussed in Sect.\ \ref{difprob.sec} give very similar results as the
top hat distributions of Sect.\ \ref{mod.sec}.  We consistently find
$\Delta \ln \tau_L < 0.18$ for the ellipticals, and $\Delta \ln \tau_L
< 0.35$ for the S0s in the outer part of the cluster.  Since $\Delta
\ln \tau_L = \Delta \tau_L / \langle \tau_L \rangle$, these numbers
are a combined constraint on the spread in ages of the galaxies, and
the mean age of the galaxy population.

\section{Evolution of the Scatter in the CM Relation}

All of the models discussed in the previous Section reproduce the
observed scatter in the CM relation of the ellipticals and S0s at
$z=0.33$.  These models make very different predictions for the
scatter in the CM relation at other redshifts. Therefore, in
principle, the models can be constrained by observations of the
scatter in the CM relation at different redshifts.  In this Section,
the predictions from the models discussed in Sect. \ref{mod.sec} are
compared to data from the literature on the central regions of the
Coma cluster at $z=0.02$, and the central regions of three clusters at
$z\sim 0.55$.  At present, there is no data available in the
literature for the scatter in the CM relation of early-type galaxies
in the outskirts of clusters, where we measure the high scatter for
the S0 galaxies.

\begin{figure*}
\begin{center}
\leavevmode 
\hbox{%
\epsfxsize=\hsize
\epsffile[18 447 592 663]{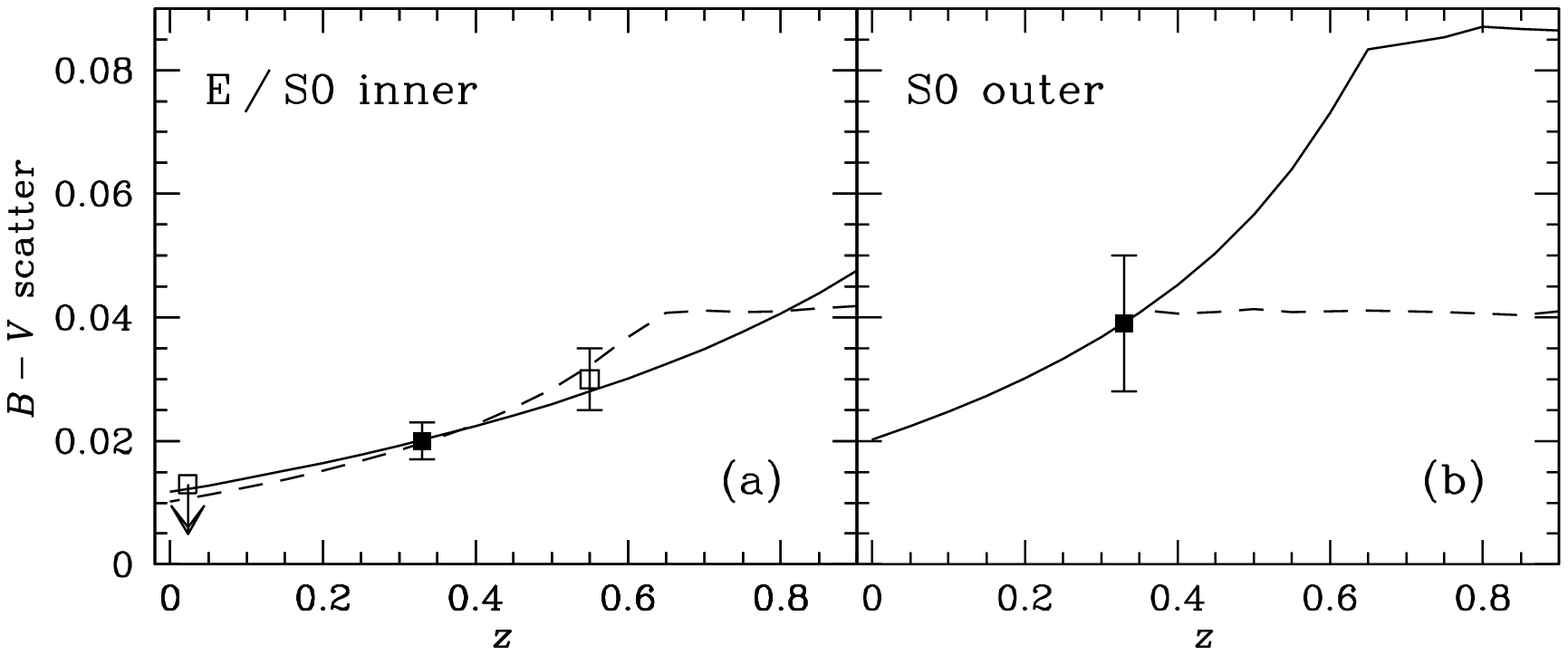}}
\begin{small}
\figcaption{\label{cmevo.plot}
The evolution of the scatter in the CM relation.
Solid symbols indicate the observed scatter of the ellipticals
in CL\,1358+62 (a), and of the S0s in the outer parts of CL\,1358+62 (b).
Open symbols in (a) represent the ellipticals in the core of the Coma
cluster at $z=0.02$ from Bower et al. (1992),
and the ellipticals and S0s in the cores of three clusters at $z=0.55$
from Ellis et al. (1997), transformed to restframe $B-V$.
The lines are the predictions from the models
discussed in Sect. \ref{mod.sec}. The models in (a) reproduce
the scatter in the CM relation
of the ellipticals in CL\,1358+62, and of the S0s in the inner parts.
The models in (b) reproduce
the scatter of the S0s in the outer parts of CL\,1358+62.
Solid lines are single burst models, broken lines are truncated continuous
star formation models (cf. Sect. \ref{mod.sec}).
In the truncated star formation model of (b), the S0s are still forming
at $z=0.33$. Therefore, the scatter in the CM relation between $z=0$
and $z=0.33$ is not well constrained in this model.
}
\end{small}
\end{center}  
\end{figure*}

Figure~\ref{cmevo.plot}(a,b) shows the predictions of the single burst
and truncated star formation models in Sect.~\ref{mod.sec} for the   
evolution of the scatter in the CM relation with redshift.
Predictions for the ellipticals and the S0s in the outer part of the
cluster are shown.  Note that the scatter is constant with time while
the galaxies were being formed.  For {\em any} continuous formation 
process the scatter is constant because it arises from $\Delta
\tau / \langle \tau \rangle$, and the mean age of the galaxies
increases at the same rate as the scatter in the ages.  After the end
of the formation phase of the galaxies, the scatter decreases
monotonically, because the relative age differences between the
galaxies become smaller.  Therefore, a nearly constant scatter in the
CM relation with redshift, such as observed by Stanford et al. (1997),
implies that the galaxies either formed at very high redshift, or
that galaxies are continuously formed and added to the sample.


For the ellipticals, it is not possible to distinguish between the
truncated star formation model and the single burst model at $z
\lesssim 0.8$.  For the S0s in the outer parts, the predictions of the
single burst model and the truncated star formation model discussed in
Sect. \ref{mod.sec} are already a factor two different at $z \geq
0.6$.  Note that the scatter at $z=0$ is not well constrained in the
truncated star formation model (the broken line in
Fig.~\ref{cmevo.plot}b). The constraint is weak because the S0s are
still forming at $z=0.33$: the scatter may be as low as $0.015$ at
$z=0$ if the production of S0s stops at $z=0.33$, or remain constant
at $\sim 0.040$ if the formation continues up to the present epoch.

The solid symbols indicate the observed scatter in the ellipticals
(Fig.~\ref{cmevo.plot}a) and the S0s in the outer parts of the cluster
(Fig.~\ref{cmevo.plot}b) at $z=0.33$.  The open symbols in
Fig. \ref{cmevo.plot}(a) are derived from data for the central regions
of the Coma cluster by Bower, Lucey, \& Ellis (1992a,b), and from the
Ellis et al. (1997) study of the CM relation in the cores of three
clusters at $z\sim 0.55$.  The Bower et al. (1992a) and Ellis et
al. (1997) data are in restframe $U-V$.  The color evolution in $U-V$
is a factor $1.5 - 2.5$ stronger than in $B-V$, depending on the
metallicity (Worthey 1994).  The $U-V$ data were transformed to $B-V$
through $\Delta U-V = 2 \Delta B-V$.

As can be seen in Fig. \ref{cmevo.plot}(a), the measurements of the   
scatter at $z=0.02$ and at $z=0.55$ are consistent with the model
predictions. The measurements we plot in Fig. \ref{cmevo.plot}(a)
between $z=0$ and $z=0.55$ are in apparent conflict with Stanford et  
al. (1997), who found that the scatter in the ``blue -- red'' (roughly
$U-V$, cf. Stanford et al. 1997) CM relation in the cores of clusters
out to $z\sim 0.9$ is approximately constant with redshift, at
$\sim 0.07$, or equivalent to $\sim 0.03$ in $B-V$.  The model
predictions range from $0.02$ at $z=0.3$ to $0.045$ at $z=0.9$.  
However, judging from Fig. 6 in Stanford et al. (1997), their error bars
are consistent with the modest increase in the scatter shown in
Fig. \ref{cmevo.plot}(a).

We cannot compare our results for the scatter in the CM relation of
the S0s in the outer part of the cluster to the above mentioned
studies, since they do not extend far enough in radius ($R \sim
0.6$\h50min\,Mpc, compared to $R = 1.6$\h50min\,Mpc for our dataset).
At present, large field photometric studies of the Coma cluster (e.g.,
Godwin et al. 1983, Mazure et al. 1988) are not sufficiently accurate
to allow a meaningful comparison with the high $z$ studies of the CM
relation.  Large field ground based studies of intermediate redshift
clusters such as that of Abraham et al.\ (1996b) lack the resolution to
determine galaxy morphologies, and sufficiently accurate colors.  Our
results, and those of Caldwell et al. (1993) for the Coma cluster,
suggest that it may be of interest to extend accurate studies of the
scatter in the CM relation of early-types to larger radii, at high and
low redshift.

\section{Infall and the Progenitors of Present-day Early-types}
\label{disc.sec}

There is good evidence that many clusters of galaxies have significant
substructure (e.g., Forman et al. 1981, Geller \& Beers 1982, Dressler
\& Shectman 1988, White, Briel, \& Henry 1993).  This is consistent
with theoretical predictions: it is expected that clusters accrete a
significant amount of mass after their initial collapse (e.g., Gunn \&
Gott 1972, Evrard 1990, 1991, Frenk et al. 1996). During the
accretion, it is likely that galaxies are transformed from star
forming galaxies into non-star forming galaxies, to produce the
passively evolving galaxies that are observed in rich clusters.  The
galaxies in which the star formation has been shut off will be
observed as blue, non-star forming galaxies for a short period ($\sim
1$\,Gyr).  These galaxies are expected to be more abundant in the
outskirts of the cluster (Evrard 1991; Frenk et al. 1996).  The blue
early-type galaxies that we observe in CL\,1358+62 may therefore be
recently accreted galaxies that have had their formation terminated
upon entry.

The accretion process may be related to the Butcher-Oemler effect
(e.g., Dressler \& Gunn 1983, Butcher \& Oemler 1984, Lavery \& Henry
1994, Moore et al. 1996, Abraham et al. 1996b).  It is often suggested
that the Butcher-Oemler effect is caused by an enhanced accretion at
higher redshifts (e.g., Kauffmann 1995).  The cluster CL\,1358+62
displays a mild Butcher-Oemler effect (Luppino et al. 1991; Fabricant
et al. 1991), and it will be interesting to see whether the scatter of
the early-type galaxies in the outer parts correlates with the
Butcher-Oemler effect.

If galaxies are accreted from the field, with a subsequent cutoff in
their star formation rate, then the set of non-star forming galaxies
in clusters evolves with time. Hence the non-star forming  
galaxies at $z=0.33$ are a subsample of the non-star forming galaxies
at $z=0$, for the same cluster.  This evolution makes it harder to  
compare samples of early-type cluster galaxies at different redshifts:
at high redshift there will be an observational bias towards the oldest
progenitors of present day early type galaxies.

We calculate the importance of this effect by estimating the number of
galaxies that will be added to the early-type galaxy population
between $z=0.33$ and $z=0$.  The subsample that is consistent with
continuous accretion (the S0s in the outer parts of the cluster)
comprises about $1/3$ of the early-type galaxy population of
CL\,1358+62. This implies that the accretion rate must have decreased
considerably since the onset of the formation of the cluster, which is
consistent with theoretical predictions (e.g., Lacey \& Cole 1993,
Kauffmann 1995).  If we assume that the accretion rate remains
constant after $z=0.33$, we obtain an upper limit on the accreted and
transformed fraction.  For $q_0 = 0.5$ we derive that only $\sim
15$\,\% of the early-type galaxy population at $z=0$ has been added
after $z=0.33$.  The main uncertainty is the limited field of view of
our data.

The transformation rate is smaller when only bright galaxies are
considered. As discussed in Sect.~\ref{bluegal.sec}, the bluest
galaxies have low luminosities. The most massive galaxies were
therefore probably transformed into passively evolving systems at
earlier times.  It is difficult to add young bright galaxies to the   
cluster sample, even when we consider mergers.  The merger of the two
brightest galaxies that are more than 0.10 magnitudes bluer than the
CM relation would produce a galaxy of $V_z \sim 19.3$.  This galaxy
would be placed among the brightest galaxies in CL\,1358+62, but only
for a short time.  The luminosity of such a blue galaxy would be
expected to fade by $\sim 0.7$ magnitudes, and the implication is that
we have not (yet) observed the young progenitors of massive galaxies.

\section{Summary and Conclusions}

We have studied the color-magnitude relation in the rich cluster
CL\,1358+62 at $z=0.33$. The elliptical galaxies in CL\,1358+62 form a
very homogeneous population: the observed scatter in the $B-V$ CM
relation is only $0.024$ magnitudes, and there is no evidence for an
increase in the scatter with distance from the cluster center. We
cannot presently exclude a small ($\lesssim 0.01$ magnitudes) trend in
the mean color of the ellipticals.

The S0s are a more heterogeneous population. The scatter and the mean
color of the CM relation are strong functions of galaxy distance from
the cluster center. The CM relations of the ellipticals and the S0s in
the inner parts of the cluster are very similar, as is the case for
three clusters at $z\sim 0.55$ studied by Ellis et al.~(1997).  The
S0s in the outer parts of CL\,1358+62 are bluer by 0.02 magnitudes,
and have a larger scatter than the S0s in the inner parts ($0.043$
vs. $0.021$).  Fig.~\ref{spatial.plot} demonstrates that the mildly
blue S0s avoid the cluster core. The morphologies of the mildly blue
S0s are very similar to those of the red S0s (cf.
Fig.~\ref{selgal.plot} [Plate 4]). The very blue S0s generally have
low luminosities.

The systematic blueing of the CM relation with radius has been
observed before (in the cluster Abell 2390 at $z=0.23$, by Abraham et
al.\ 1996b), but this is the first time that the morphologies of the
galaxies at large distances from the cluster center could be
determined. Also, the colors in the study of Abraham et al.\ (1996b)
are not accurate enough for the detection of the radial gradient in
the scatter in the color-magnitude relation.

We have constructed a range of models to explain the scatter in the CM
relation of the ellipticals and the S0s in the outer parts of the
cluster. We find that truncated continuous star formation models allow
for more recent star formation than single burst models.  The youngest
ellipticals formed before $z = 1.2$ if they formed in single bursts of
star formation, and before $z=0.6$ if they experienced a constant star
formation rate until an abrupt truncation.  In all models, the
youngest S0s in the outer parts of the cluster formed stars until
close to the epoch of observation. In the truncated star formation
model, star formation in these galaxies continued until the epoch of
observation.


The radial trend of the S0 colors is consistent with the hypothesis
that galaxies and groups are continuously accreted from the field. In
this picture, the blue S0s are the galaxies that have been accreted
recently.  The star formation in the infalling galaxies is turned off
upon entering the cluster. The truncation of the star formation may be
accompanied by a small starburst in the nucleus (e.g., Hernquist \&
Mihos 1995). This nuclear starburst would explain the blue bulges (and
the small color gradients) that we see in many of the blue S0 galaxies
(cf. Sect. \ref{colgrad.sec}).  The accretion radius of the cluster is
roughly $3$\,\h50min\,Mpc (Carlberg, Yee \& Ellingson 1997), which is
just outside our HST mosaic.  If the morphological transformation
between spirals and S0s occurs before accretion into the denser
regions of the cluster, while the galaxies are still in small groups,
that process probably occurs at radii beyond the limit of our
observations.  This might explain the small fraction of spirals in
this very rich cluster.

The continuous infall of field galaxies implies that the fraction of 
non-star forming galaxies in clusters evolves with time. This  
complicates the comparison between galaxy populations in high redshift
clusters and local clusters. Assuming a constant accretion rate after 
$z=0.33$, we estimate that $\sim 15$\,\% of the present-day early-type   
galaxy population in clusters has been added between $z=0.33$ and   
$z=0$. We infer that the accretion rate was higher before $z=0.33$,
which is consistent with model predictions (e.g., Kauffmann 1995). 

The infall and transformation process seems to affect mostly the faint
end of the galaxy population: the bluest galaxies have low
luminosities. There are a few bright early-types with slightly bluer
colors than the CM relation.
Their presence suggests that bright early-type
galaxies with young populations may be found in clusters at higher
redshifts.

The homogeneity of the ellipticals and luminous S0s suggests that
their star formation had ceased before they were accreted onto the
cluster. There are several possible mechanisms, e.g., the star
formation may have ceased after the collapse of smaller subclumps
(galaxy groups) in which the galaxies resided. Studies of early-type
galaxies in the field will be valuable to determine the process that
shuts off the star formation in bright early-types before they enter
the cluster environment.

This study demonstrates the need for large field, high resolution
observations of distant clusters.  If only one HST WFPC2 pointing had
been available, our conclusions would be quite different: inside $R =
80''$, there is no systematic difference in the scatter or offset of
the CM relation for the ellipticals and the S0s.  Additional wide
field observations are needed to establish whether CL\,1358+62 is
typical of intermediate $z$ clusters and to extend this study to 
higher redshifts. High quality data for nearby clusters extending over
a wide field would also be valuable.  It will be interesting to      
explore whether the differences between ellipticals and S0s, and the
environmental dependence of the colors, are reflected in the
Fundamental Plane of this cluster: the observed scatter of $0.043$ in
the S0 colors in the outer parts implies a scatter in the Fundamental
Plane in the $V$ band of $\sim 30$\,\% in $\mu_{\rm e}$.

\acknowledgements{We thank the referee, Dr. Ian Smail, for his
constructive comments, which improved the paper. Henk Hoekstra
is thanked for a critical reading of the appendix. The University of
Groningen and the Leids Kerkhoven-Bosscha Fonds are thanked for support.
Support from STScI grants
GO05989.01-94A, GO05991.01-94A, and AR05798.01-94A is gratefully
acknowledged.}

\begin{appendix}
\section{Models for the color evolution}
\label{models.sec}
\subsection{Single bursts}
In this model the galaxies formed in single bursts of star formation
at $t=t_{0}$.
The luminosity evolution of a single age stellar population can
be described by a power law:
\begin{equation}
L \propto \frac{1}{(t-t_{0})^{\kappa}},
\end{equation}
with $t$ the age of the universe at the epoch of observation, and
$t_{0}$ the age of the universe at the time of formation
of the population. The difference $t-t_{0}$ is the age of
the population at the epoch of observation.
The coefficient $\kappa$ depends on the passband, but also on
the metallicity $Z$ and the IMF (Tinsley 1980; Worthey 1994).

The color evolution is
\begin{equation}
\label{bvsi.eq}
\frac{L_V}{L_B} \propto (t-t_{0})^{\kappa_B-\kappa_V},
\end{equation}
or
\begin{equation}
\label{bvs.eq}
B-V = 1.086 (\kappa_B - \kappa_V) \ln (t-t_{0}) + C,
\end{equation}
where we have set $B-V \equiv 2.5 \log (L_V/L_B)$.
The evolution of the color depends
on the difference $\kappa_B - \kappa_V$, which is not
very sensitive to differences in $Z$ or the IMF. The Worthey
(1994) models give $\kappa_B - \kappa_V \approx 0.10$
for $-2 < Z < 0.5$. 
In the single burst model, the scatter in the colors at time
$t$ is proportional to the scatter in $\ln (t-t_{0})$.

\subsection{Secondary bursts}
This model is an extension of the single burst model. A mass fraction
$(1-f)$ forms in a burst at $t=t_{0}$, and a mass fraction $f$
forms in a burst at $t=t_{1}$, with $t_{1} > t_{0}$.
The luminosity evolves as
\begin{equation}
L \propto \frac{1-f}{(t-t_{0})^{\kappa}} + \frac{f}{(t-t_{1})
^{\kappa}},
\end{equation}
and the color as
\begin{equation}
\label{goodsec.eq}
\frac{L_V}{L_B} \propto { (1-f)(t-t_{0})^{-\kappa_V} +
f (t-t_{1})^{-\kappa_V}  \over (1-f)(t-t_{0})^{-\kappa_B} +
f (t-t_{1})^{-\kappa_B} }.
\end{equation}
Expressed in magnitudes this is
\begin{equation}
\label{bvb.eq}
B - V \approx 1.086 (\kappa_B - \kappa_V) \ln (t-t_{0}) -
1.086 \frac{f}{1-f} \left[ \left( \frac{t-t_{1}}{t-t_{0}} \right)
^{-\kappa_B} - \left( \frac{t-t_{1}}{t-t_{0}} \right)
^{-\kappa_V} \right] + C,
\end{equation}
where it is assumed that $f \ll \frac{t-t_1}{ t-t_{0} }$, i.e.,
Eq.~\ref{bvb.eq} is not valid shortly after a strong burst.
In that case, the color evolution reduces to the single burst case
(Eq.~\ref{bvs.eq}), with $t_{0}$ replaced by $t_{1}$.

The observed scatter in the colors can be expressed as a
constraint on the relevance of star bursts. It is assumed the scatter
in the colors is caused by a spread in ages of the bursts only (i.e.,
not by a spread in $t_{0}$ or $f$).
Differentiating Eq.~\ref{bvb.eq} with respect to $(t-t_{1})$
gives an expression for the scatter in the colors as a function of
the scatter in $(t-t_{1})$:
\begin{equation}
\label{dbvbf.eq}
\Delta(B-V) = 1.086 \frac{f}{1-f} \left[ \kappa_B
\left( \frac{t-t_{1}}{t-t_{0}} \right)^{-\kappa_B - 1} -
\kappa_V
\left( \frac{t-t_{1}}{t-t_{0}} \right)^{-\kappa_V - 1}
\right] \frac{\Delta (t-t_1)}{t-t_{0}}.
\end{equation}
For $1.5 < (\kappa_B + \kappa_V) < 2$,
a good approximation to Eq.~\ref{dbvbf.eq} is
\begin{equation}
\label{dbvb.eq}
\Delta (B-V) \approx
1.086 (\kappa_B - \kappa_V) \frac{f}{1-f}\left(\frac{t-t_0}{t-t_1}
\right)^{1.5}
\Delta \ln (t-t_1).
\end{equation}

\subsection{Truncated star formation}
In this model the galaxies form stars at a constant rate from $t=t_0$
to $t=t_{1}$. The star formation is truncated at $t=t_{1}$.
This model is appropriate for spiral galaxies in which the
star formation suddenly terminates.
The ages of the stellar populations in the galaxies range from
$(t-t_{0})$ to $(t-t_{1})$.
The contribution to the luminosity
of each population is $dt_* (t-t_*)^{-\kappa}$.
Therefore, the total luminosity after $t=t_1$ is
\begin{equation}
L \propto \frac{1}{t_{1} - t_0} \int_{t_0}^{t_{1}}
\frac{dt_*}{(t-t_*)^{\kappa}}.
\end{equation}
The factor $1/(t_{1} - t_0)$ normalizes the mass such that
the mass of the galaxy after $t_{1}$ is independent of the
length of the burst. For $\kappa_B$, $\kappa_V \neq 1$
the color evolution is
\begin{eqnarray}
\frac{L_V}{L_B} &\propto& \frac{1-\kappa_B}{1-\kappa_V} \left[ {
(t-t_0)^{1-\kappa_V} - (t-t_{1})^{1-\kappa_V}  \over
(t-t_0)^{1-\kappa_B} - (t-t_{1})^{1-\kappa_B} } \right]\\
&=&
\frac{1-\kappa_B}{1-\kappa_V} (t-t_0)^{\kappa_B - \kappa_V}
\left( \frac{t-t_{1}}{t-t_0} \right) ^{\frac{1}{2}
(\kappa_B-\kappa_V)}
\frac{\left( \frac{t-t_{1}}{t-t_0} \right) ^{-\frac{1}{2}
(1-\kappa_V)} - \left( \frac{t-t_{1}}{t-t_0} \right) ^{\frac{1}{2}
(1-\kappa_V)}}{
\left( \frac{t-t_{1}}{t-t_0} \right) ^{-\frac{1}{2}
(1-\kappa_B)} - \left( \frac{t-t_{1}}{t-t_0} \right) ^{\frac{1}{2} 
(1-\kappa_B)}}. \nonumber
\end{eqnarray}
Since
\begin{equation} 
\frac{x^{-\alpha} - x^{\alpha}}{x^{-\beta} - x^{\beta}} =
\frac{ \alpha \log x + \sum_{k=2}^{\infty} \frac{1}{(2k-1)!}
 (\alpha \log x)^{2k-1} }
{\beta \log x + \sum_{k=2}^{\infty} \frac{1}{(2k-1)!}
 (\beta \log x)^{2k-1} }
\approx \frac{\alpha}{\beta},\nonumber
\end{equation}
the color evolution reduces to
\begin{equation}
\label{bvti.eq}
\frac{L_V}{L_B} \propto
\left[\sqrt{(t-t_0) (t-t_{1})}\right]^{\kappa_B - \kappa_V}.
\end{equation}
In the truncated star formation model, galaxies are comprised of
populations spanning a range of ages.
The expression for the color evolution in this model is very similar
to that for the color evolution in the single burst model
(Eq.~\ref{bvsi.eq}).
However, the rate of evolution is determined by
the geometric mean of the ages of the youngest stars and the oldest
stars in the galaxy, rather than
a single age.\\
Expressed in magnitudes, Eq.~\ref{bvti.eq} reads
\begin{equation}
\label{bvt.eq}
B-V = 0.543 (\kappa_B - \kappa_V)
\left[\, \ln (t-t_0) + \ln (t-t_{1}) \right] + C.
\end{equation}
If it is assumed all galaxies have the same $t_0$,
the scatter in the $B-V$ colors at time $t$ only depends on the
scatter in truncation times $t_{1}$.

\subsection{Mean ages}

Here, we give expressions for
the luminosity weighted mean ages $\tau_L$ in the three models.
In the single burst model,
all the stars have the same age, and therefore
\begin{equation}
\tau_L = t-t_0.
\end{equation}
In the secondary burst model
\begin{equation}
\tau_L = \frac{(1-f) (t-t_0)^{1-\kappa} +
f (t-t_1)^{1-\kappa}}{(1-f) (t-t_0)^{-\kappa} +
f (t-t_1)^{-\kappa}},
\end{equation}
and in the truncated star formation model
\begin{equation}
\tau_L = \frac{1-\kappa}{2-\kappa}
\left[ \frac{(t-t_0)^{2-\kappa} - (t-t_1)^{2-\kappa}}
{(t-t_0)^{1-\kappa} - (t-t_1)^{1-\kappa}} \right].
\end{equation}

\section{Correction for Luminosity Evolution}

If the scatter in the CM relation is caused by age variations
among the galaxies,
the observed scatter in the CM relation is partly caused by
the increased luminosity of young galaxies.
The deviation for a given object in $B-V$ from the CM relation can
be expressed as
\begin{equation}
\Delta (B-V)_{\rm obs} = \Delta (B-V)_{\rm evo} - \alpha \Delta V,
\end{equation}
where $\alpha$ is the slope of the CM relation and
$\Delta V$ is the amount of luminosity evolution for a color evolution
of $\Delta (B-V)_{\rm evo}$. Since the relation between color evolution
and luminosity evolution is
\begin{equation}
\Delta (B-V)_{\rm evo} = 1.086 \left( \frac{\kappa_B}{\kappa_V} -1 \right)
\Delta V,
\end{equation}
with $\kappa_B$, $\kappa_V$ defined in Appendix A, it follows
that
\begin{equation}
\Delta (B-V)_{\rm obs} = \left( 1 + \frac{\alpha}{1.086}
\frac{\kappa_V}{\kappa_V - \kappa_B} \right) \Delta (B-V)_{\rm evo}.
\end{equation}
The slope of the CM relation $\alpha = -0.018$ (see Eq.~\ref{cm.eq}).
The Worthey (1994) models give
$\kappa_B = 0.90$ and $\kappa_V = 0.80$ for solar metallicity.
These values give a  correction for the luminosity evolution of
$\Delta (B-V)_{\rm evo} \approx 0.90 \Delta (B-V)_{\rm obs}$.

\end{appendix}

\newpage

\begin{table}[t]
\begin{center}
{ {\sc TABLE 1} \\
\sc Galaxy Catalog} \\
\vspace{0.1cm}
\begin{tabular}{rrrccccccc}
\hline
\hline
Id & $x$\,($''$) & $y$\,($''$) & type & $n$ & $F814W$ & $V_z$ & $F606W-F814W$ & $(B-V)_z$ & $r_{\rm c}$\,($''$) \\
\hline
  86 & $ -60.9 $ & $ -252.6 $ &      E & 1 & 20.78 &  21.68 & 1.179 & 0.837 & 0.32 \\ 
  92 & $ 137.2 $ & $ -237.1 $ &     S0 & 2 & 19.84 &  20.75 & 1.201 & 0.855 & 0.60 \\ 
  95 & $ 63.5 $ & $ -235.4 $ &      E & 3 & 19.32 &  20.25 & 1.249 & 0.894 & 0.52 \\ 
\hline
\tablecomments{Table 1 is published in its entirety in the AAS CD-ROM Series.}
\end{tabular}
\end{center}
\end{table}

\begin{table}[h]
\begin{center}
{ {\sc TABLE 2} \\
\sc Offset and Scatter of the CM Relation} \\
\vspace{0.1cm}
\begin{tabular}{lrrcccc}
\hline
\hline
Sample & $N$ & Offset & Error & Obs.\ Scatter & Intr.\ Scatter & Error \\
\hline
All & 188 & $ -0.012 $ & 0.003 & 0.041 & 0.040 & 0.006 \\ 
E& 46 & $ 0.000 $ & 0.003 & 0.024 & 0.022 & 0.003 \\ 
S0& 95 & $ -0.012 $ & 0.003 & 0.031 & 0.029 & 0.004 \\ 
S &  7 & $ -0.141 $ & 0.032 & 0.092 & 0.091 & 0.051 \\ 
Irr&  9 & $ -0.292 $ & 0.027 & 0.086 & 0.085 & 0.023 \\ 
$n=4$& 46 & $ -0.008 $ & 0.003 & 0.024 & 0.022 & 0.005 \\ 
$n=3$&  80 & $ -0.011 $ & 0.004 & 0.036 & 0.034 & 0.006 \\ 
$n=2$&  32 & $ -0.022 $ & 0.010 & 0.061 & 0.060 & 0.026 \\ 
$n=1$&  30 & $ -0.117 $ & 0.026 & 0.154 & 0.154 & 0.021 \\ 
E $<118''$& 24 & $ 0.003 $ & 0.005 & 0.024 & 0.021 & 0.004 \\ 
E $\geq 118''$& 22 & $ -0.003 $ & 0.005 & 0.025 & 0.023 & 0.006 \\ 
S0 $<118''$& 46 & $ -0.005 $ & 0.003 & 0.021 & 0.018 & 0.004 \\ 
S0 $\geq 118''$& 49 & $ -0.019 $ & 0.006 & 0.043 & 0.041 & 0.009 \\ 
\hline
\end{tabular}
\end{center}
\end{table}

\setcounter{figure}{0}

\figcaption{\label{mosaic.plot}{\it Hubble Space Telescope} WFPC2 mosaic
of the cluster CL\,1358+62 at $z=0.33$. The image is a mosaic of
twelve adjacent HST pointings, in two filters ($F606W$ and
$F814W$). The $F606W$ and $F814W$ images were added to increase the
S/N, giving a total exposure time of 7200\,s per pointing (3600\,s in each
filter).  North is up and East is to the left.  The length of the
scalebar is 1 arcmin, or 350\,\h50min\,kpc at the distance of
CL\,1358+62 ($q_0 = 0.5$). The total area of the image is
49\,arcmin$^2$. The galaxy density decreases with distance from
the cluster center and BCG. The galaxy population in this intermediate
$z$ cluster can be studied in lower density regions than in previous
studies with HST, which focussed on the cluster cores.}

\figcaption{\label{center.plot}Color representation of the central part
of Fig. \ref{mosaic.plot} [Plate 1], created from the $F606W$ and
$F814W$ exposures. The cluster members are easily recognized by their
yellow colors. However, many of the bluer galaxies are also cluster
members.}

\figcaption{\label{allgal.plot}Greyscale representations of all spectroscopically
confirmed cluster members in the CL\,1358+62 HST mosaic.
The boxes are $6\farcs4$, or $37$\,\h50min\,kpc ($q_0 = 0.5$), on a side.}

\figcaption{\label{selgal.plot}Examples of early-type galaxies on the CM
relation, examples of slightly blue early-types, all very blue
early-types,
all spirals, and examples of irregular galaxies.  Each box is
$6\farcs4 \times 6\farcs4$ ($37 \times 37$\,\h50min\,kpc).
The number in the lower right of each box
is the galaxy identification, the number in the lower left is the
restframe $B-V$ color with respect to the CM relation.  The slightly
blue early-type galaxies have very similar morphologies to the
early-types on the CM relation. The very blue early-types have
low luminosities and generally have significant disks.}

\end{document}